\documentclass[epj,final]{svjour} 
\usepackage{graphicx}% Include figure files 
\usepackage{dcolumn}% Align table columns on decimal point 
\usepackage{bm}% bold math 
\usepackage{psboxit,epsfig,wrapfig,boxedminipage,enumerate} 
\usepackage{nicefrac} 
\usepackage{color} 
\usepackage{rotating} 
\usepackage{amsmath} 
\usepackage{amssymb} 
 
\newcommand{\bv}{{\bf v}} 
\newcommand{\br}{{\bf r}} 
\newcommand{\bk}{{\bf k}} 
\newcommand{\bF}{{\bf F}} 
\newcommand{\bforce}{{\bf f}}

\def\dblone{\hbox{$1 \hskip -1.2pt\vrule depth 0pt height 1.6ex width 
    0.7pt \vrule depth 0pt height 0.3pt width 0.12em$}}

\begin{document} 
 
\title{Collective dynamics of colloids at fluid interfaces} 
 
\author{J.~Bleibel\inst{1,2}, A.~Dom\'\i nguez\inst{3}, M.~Oettel\inst{2}, 
  S.~Dietrich\inst{1,4}}   
\institute{Max-Planck-Institut f\"ur Intelligente Systeme, 
  Heisenbergstr.~3, 70569 Stuttgart, Germany \and  
  Institut f\"ur Physik, WA 331, Johannes Gutenberg 
  Universit\"at Mainz, 55099 Mainz, Germany \and 
  F\'\i sica Te\'orica, Universidad de Sevilla, Apdo.~1065, 
  41080 Sevilla, Spain \and 
  Institut f\"ur Theoretische und Angewandte Physik, 
  Universit\"at Stuttgart, Pfaffenwaldring 57, 70569 Stuttgart, Germany} 
 
\date{\today} 
 
\abstract{ 
  The evolution of an initially prepared distribution of micron sized colloidal 
  particles, trapped at a fluid interface and under the 
  action of their mutual capillary attraction, is analyzed by using 
  Brownian dynamics simulations. At a separation $\lambda$ given by the 
  capillary length of typically $1\; \rm mm$, the distance dependence of this 
  attraction exhibits a crossover from a logarithmic decay, formally 
  analogous to two--dimensional gravity, to an exponential decay. We discuss in 
  detail the adaption of a particle--mesh algorithm, as used in cosmological 
  simulations to study structure formation due to gravitational collapse, to 
  the present colloidal problem. These simulations confirm the predictions, 
  as far as available, of a mean--field theory developed previously for 
  this problem.   
  The evolution is monitored by quantitative characteristics which  
  are particularly sensitive to the formation of highly inhomogeneous 
  structures. Upon increasing $\lambda$ the dynamics show a smooth transition 
  from the spinodal decomposition expected for a simple fluid with 
  short--ranged attraction to the self--gravitational collapse scenario.}


\maketitle 
 
%%%%%%%%%%%%%%%%%%%%%%%%%%%%%%%%%%%%%%%%%%%%%%%%%%%%%%%%%%%%%%%% 
%%%%%%%%%%%%%%%%%%%%%%%%%%%%%%%%%%%%%%%%%%%%%%%%%%%%%%%%%%%%%%%% 
 
\section{Introduction} %: 2D Cosmology in a petri dish} 
\label{intro} 
 
The investigation of systems in low dimensions offers a rich spectrum of 
phenomena which are rather distinct from three--dimensional (3D) bulk 
properties. This is the case because structural properties of condensed matter 
depend sensitively on the spatial dimension. 
For example, in two--dimensional (2D) systems the solid--liquid phase 
transition is replaced by a solid--hexatic--liquid transition 
\cite{Kosterlitz:1973,Bladon:1995,Marcus:1997}.  
Micron sized colloidal particles with radii $R_0$ trapped at a fluid interface 
form a monolayer and 
constitute an easily accessible mesoscopic model system which exhibits many 
characteristics of truly 2D systems. This system can actually have  
fluid and crystalline phases~\cite{Pieranski:1980},  
as well as a hexatic phase~\cite{ZLM99}. 
Applications may involve 
controlled self assembly~\cite{Bowden:1997} and structure 
formation~\cite{Aizenberg:2000,Helseth:2005,Loudet:2009}, exploiting the 
specific features of the interactions between colloidal 
particles.  
 
\begin{figure}[h] 
  \begin{minipage}{0.20\linewidth} 
    \epsfig{file=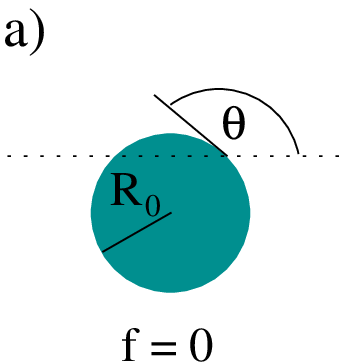,width=\linewidth} 
  \end{minipage} 
  \begin{minipage}{0.785\linewidth} 
    \epsfig{file=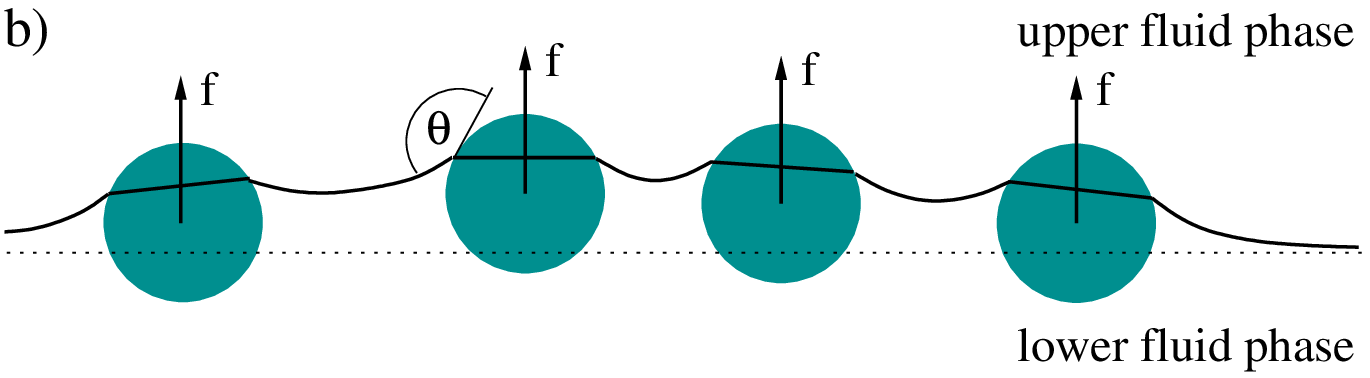,width=\linewidth} 
  \end{minipage} 
  \caption{\label{fig_1} a) A single colloid adsorbed at a fluid interface 
    (dotted line) with no external forces present. b) Identical particles 
    trapped at the interface between two fluids and deforming it due to an 
    external force $f$ acting on each of them~\cite{Dominguez:2010} Upwards 
    pointing forces are taken to be positive. In all cases the interface 
      meets the surfaces of the colloids under a fixed contact angle 
      $\Theta$.} 
\end{figure}

Specifically,  
the focus of the present study is 2D particle aggregation due to 
interactions induced by interfacial deformation, i.e.,  
capillary attraction (see, e.g., Refs.~\cite{Oettel:2008,Domi10} and 
references therein). 
We consider a collection of 
colloidal particles at the interface between two fluid phases, each colloid 
being acted upon vertically by  
an external force $f$ (due to, e.g., gravity, an external electric 
field, an optical tweezer, etc.; see Fig.~\ref{fig_1} and 
  Ref.~\cite{Dominguez:2010}).  
  
In the limit 
of small interfacial deformations or, alternatively, if the 
interparticle separation $d$ between the centers of the colloids is large 
enough,  
the effective capillary attraction is described by the 
pair potential  
\cite{Nico49,CHW81,PKDN93,ODD05,DOD05,Mueller:2005,Dominguez:2008}  
\begin{equation} 
  \label{eq2} 
  V_\mathrm{cap}(d) = - V_0 K_0 \left(\frac{d}{\lambda}\right) , 
  \qquad 
  V_0 := \frac{f^2}{2\pi\gamma} \,; 
\end{equation} 
$\lambda:=\sqrt{\gamma/(g\Delta \rho)}$ is the \textit{capillary 
  length}, which determines the range of the interaction and is given 
in terms of the gravitational acceleration $g$, the surface tension $\gamma$ of 
the interface, and the mass density difference $\Delta \rho$ 
between the two fluid phases. The prefactor $V_0$ sets the strength of the 
interaction and $K_0$ denotes a modified Bessel function. Due to $V_{\rm 
  cap}(d \ll \lambda) \sim \ln d$, in that range the potential is 
non-integrable (in the sense of equilibrium statistical mechanics, i.e., 
leading to a hyperextensive scaling of the energy), while for $\lambda$ finite  
$V_{\rm  cap}(d \gg \lambda) \sim \exp(-d/\lambda)$. This functional 
dependence is analogous to screened electrostatics or gravity in 2D, 
with $\lambda$ playing the role of the screening length and $f$ the 
role of the electric charge or the gravitational mass. The difference with 
electrostatics is that here equal (different) charges attract (repel) 
each other, and the difference with gravity is that here one can have negative 
as well as positive masses. 
Typically, $\lambda$ is of the order of a few millimeters so that  
$\lambda\gg R_0$. This implies that a single particle interacts with many 
neighbors through a non-integrable potential and, in this sense and range, the 
capillary attraction formally qualifies as a long--ranged interaction. 
Long--ranged interactions enjoy significant current interest  
(see, e.g., Ref.~\cite{CDR09} and references therein) in many fields, such as 
self--gravitating fluids in cosmology, 
two--dimensional vortex flow in hydrodynamics, and bacterial 
chemotaxis in biology~\cite{Keller:1970,ChSi08}; 
for the present study the theoretical background reviewed in 
Ref.~\cite{Chavanis:2010} is of particular interest. 

In Ref.~\cite{Dominguez:2010} the 
instability of an initially homogeneous particle distribution, induced 
by capillary attraction, has been analyzed.  
The instability is analogous to Jeans' instability of a 
self--gravitating fluid~\cite{Jeans:1902,BiTr08}, except for the 
presence of the screening length $\lambda$.  
Since the parameters $f$, $\lambda$, $\gamma$, $R_0$ as well as state 
variables like particle density and temperature are easily 
tunable, the range of values of the parameters for 
the observation of this instability for realistic experimental 
configurations has been explored in Ref.~\cite{Dominguez:2010}. Even though 
the capillary attraction between two isolated particles can be relatively weak 
compared with other effects like thermal agitation and short--ranged repulsive 
forces, its  
long--ranged nature (in the sense $\lambda\gg R_0$) leads to a 
cumulative effect and to an eventual predominance of the capillary 
force at large scales~\cite{Bleibel:2011}. Thus, these colloidal
monolayers turn out to be   
excellent model systems for the experimental study of the collective 
evolution driven by the pair potential given in Eq.~(\ref{eq2}). We are 
aware, however, of only one systematic experimental study of this instability 
with colloidal monolayers \cite{VAKZ01}, in which the long--ranged 
capillary interactions have been identified as the driving forces of 
particle aggregation.   
 
Here we perform numerical simulations in order to 
extend the results reported in Ref.~\cite{Dominguez:2010} by going beyond the 
simplifying assumptions underlying the theoretical analysis presented there. 
The paper is organized as follows. In Sec.~\ref{sec:theory} we briefly 
recall the main theoretical results obtained in Ref.~\cite{Dominguez:2010}, 
followed in Sec.~\ref{sec:simulations} by a description of the 
algorithm of the simulation used here. This includes in particular a detailed 
account of how the long--ranged nature of the capillary interaction is 
handled, similar to the approach for simulations of cosmological 
structure formation. 
Section \ref{sec:results} presents the analysis of the simulations, 
both for the evolution of an unstable homogeneous 
distribution and the evolution of a radially symmetric particle 
density.  
Particular interest is paid to the dependence of the results on the 
ratio between the capillary length $\lambda$ and the size $L$ of the 
system. This allows one to track the crossover between non-integrability 
($R_0, L \ll \lambda$) and screening ($R_0 \ll \lambda \ll L$). 
Our conclusions are discussed and summarized in 
Sec.~\ref{sec:summary}. 
 
%%%%%%%%%%%%%%%%%%%%%%%%%%%%%%%%%%%%%%%%%%%%%%%%%%%%%%%%%%%%%%%% 
%%%%%%%%%%%%%%%%%%%%%%%%%%%%%%%%%%%%%%%%%%%%%%%%%%%%%%%%%%%%%%%% 
 
\section{Theoretical model} 
\label{sec:theory} 
 
In this section we briefly describe the theoretical model introduced in 
Ref.~\cite{Dominguez:2010}, which serves as the basis for the simulations 
presented here. 
 
Upon projection onto the reference plane, for a contact angle of 
$\Theta=90^{\circ}$,  
the configuration of spherical particles of radius $R_0$ is mapped onto a 
configuration of monodisperse discs of radius $R_0$\footnote{The 
    actual deformation of the interface due to external forces $f$ acting on 
    a colloidal particle is small compared with its radius 
    $R_0$~\cite{ODD05}. Therefore, for a contact angle of $\Theta=90^{\circ}$, 
    the discs have approximately the same radius as the spherical colloids.},  
the centers of which form an  
areal number density 
$\varrho(\br=(x,y),t)$, which evolves as a function of time under the action 
of two types of (effective) forces: the capillary force  
and a short--ranged repulsive force   
such as hard core or electrostatic repulsion. Within this model, the dynamical 
evolution of $\varrho(\br,t)$ is obtained under the following simplifying 
assumptions. 
 
$\bullet$\textit{Local equilibrium}: The characteristic time scales of 
evolution are sufficiently large so that the forces  
can be described by a local equilibrium ansatz. For the short--ranged 
forces the force per unit area is ($\nabla=(\partial_x,\partial_y)$) 
\begin{equation} 
  \label{eq:short} 
  \bF_\mathrm{short}(\br,t) \approx - \nabla p(\varrho=\varrho(\br,t)) , 
\end{equation} 
expressed in terms of the equilibrium equation of state for the pressure 
$p(\varrho)$ of the 2D gas of discs in the absence of capillary deformations.  
Isothermal conditions are assumed. The temperature is a constant parameter  
fixed by the upper and lower fluids acting as thermal baths, and thus it is 
irrelevant for the temporal evolution. 
 
$\bullet$\textit{Mean field}: The capillary force is computed assuming that  
the interfacial deformation is small and adjusts instantaneously to the 
momentary particle  
distribution, i.e., it satisfies the Young--Laplace equation for a given 
particle distribution $\varrho(\br,t)$. Furthermore, a mean--field 
approximation is used so that the capillary force per unit area can be 
expressed as   
\begin{equation} 
  \label{eq:Fcap} 
  \bF_\mathrm{cap} = f \varrho \nabla U , 
\end{equation} 
with the ensemble averaged mean--field interfacial deformation $U(\br)$.  
By using $U(\br)$ one neglects spatial variations on small scales, because it 
is determined self--consistently by the mean--field Young--Laplace 
equation: 
\begin{equation} 
  \label{eq:YL} 
  \nabla^2 U - \frac{U}{\lambda^2} = - \frac{f}{\gamma} \varrho . 
\end{equation} 
Accordingly it is computed from the \textit{average} density field 
$\varrho(\br,t)$ rather than from the density field \textit{conditional} to 
the presence of a particle at $\br$~\cite{Dominguez:2010}.  
The pair potential in Eq.~(\ref{eq2}) can be expressed as 
$V_\mathrm{cap}(d) = f U_\delta(d)$ in terms of Green's 
function $U_\delta(d)$ of Eq.~(\ref{eq:YL}), i.e., the solution for a 
single--particle source $\varrho(\br) = \delta(\br)$. 
This approximation is justified by   
the long range $\sim 
\lambda$ (see Eq.~(\ref{eq2})) of the capillary force: if $\varrho_\mathrm{h}$ is 
a characteristic homogeneous particle number density, the number of particles 
with which a single particle interacts is effectively of the order 
$\varrho_\mathrm{h} \lambda^2 \sim (\lambda/R_0)^2 \gg 1$.  
 
$\bullet$\textit{Overdamped motion}: Inertia of the particles is 
neglected, 
so that the in-plane velocity field $\bv(\br, t)$ of the 2D particle 
distribution is proportional to the driving force, 
\begin{equation} 
  \label{eq:flow} 
  \varrho \bv(\br, t) = \Gamma ( \bF_\mathrm{short} + \bF_\mathrm{cap} ) , 
\end{equation} 
where $\Gamma$, in units of time per mass, is the effective single--particle 
mobility at the interface. For reasons of simplicity in Eq.~(\ref{eq:flow}) 
the effect of hydrodynamic interactions is neglected and $\Gamma$ is taken to be 
constant in space and time\footnote{Hydrodynamic interactions play an 
  important role in the process of particle trapping by the interface, 
  as studied recently in Ref.~\cite{Singh:2010}. However, the present theory 
  only considers the evolution \emph{in} the plane of the   
  interface \emph{after} the particles have been trapped there and equilibrium 
  with respect to their distribution normal to the interface has  
  been reached.}. 
 
Under these conditions, mass conservation leads to the following form 
of the continuity equation: 
\begin{equation} 
  \label{eq:continuity} 
  \frac{\partial \varrho}{\partial t} = - \nabla\cdot(\varrho \bv) =  
  \Gamma \nabla\cdot ( \nabla p - f \varrho \nabla U) . 
\end{equation} 
Qualitatively, this equation describes the competition between 
clustering driven by the capillary attraction and the homogenization 
tendency caused by the short--ranged forces. 
The solutions of this equation have been studied theoretically in 
Ref.~\cite{Dominguez:2010} under certain simplifying conditions. The 
corresponding findings can be summarized as follows: 
 
$\bullet$ \textit{Jeans' instability}: 
With the  homogeneous density field $\varrho_\mathrm{h}$ one can  
associate Jeans' time,  
\begin{equation} 
  \label{eq:jeansT} 
  \mathcal{T}=\frac{\gamma}{\Gamma f^2 \varrho_\mathrm{h}} , 
\end{equation} 
and Jeans' wavenumber, 
\begin{equation} 
  \label{eq:jeansK} 
  K = \sqrt{\frac{f^2 \varrho_\mathrm{h}^2 \kappa_\mathrm{h}}{\gamma}} , 
\end{equation} 
where $\kappa_\mathrm{h} = (\varrho_\mathrm{h} p'(\varrho_\mathrm{h}))^{-1}$ 
is the isothermal compressibility corresponding to the equation of state 
$p(\varrho)$ for a homogeneous configuration.  
A linear stability analysis shows that a 
small sinusoidal perturbation $\varrho_\bk(\br, 0) = C \varrho_\mathrm{h} 
\exp{( - i \bk\cdot\br)}$, $|C|\ll 1$, of the initially homogeneous 
distribution $\varrho_\mathrm{h}$ evolves in time as 
\begin{equation} 
  \label{eq:linearEvol} 
  \varrho_\bk(\br, t) = C \varrho_\mathrm{h} \exp{( - i 
  \bk\cdot\br + t/\tau(k))} , 
\end{equation} 
with a wavenumber-dependent characteristic time 
\begin{equation} 
  \label{eq:tau} 
  \tau(k) = \mathcal{T} \left(\frac{K}{k}\right)^2 
  \left[ \frac{1}{(k/K)^2 + (\lambda K)^{-2}} - 1 \right]^{-1} . 
\end{equation} 
From this expression one infers that long--wavelength  
perturbations with 
\begin{equation} 
  \label{eq:Kc} 
  k < K_\mathrm{c} := K \sqrt{1 - \frac{1}{(\lambda K)^2}}  
\end{equation} 
grow in time, i.e., $\tau(k) > 0$. Thus the homogeneous state is 
unstable, provided $\lambda K > 1$.  
This is known as Jeans' instability in a 3D  
self--gravitating gas  
\cite{Jeans:1902,Kiessling:1999eq}.  
 
$\bullet$ \textit{Cold collapse:} This approximation amounts to 
neglecting $\bF_\mathrm{short}$ in Eq.~(\ref{eq:flow}) relative to 
the capillary force. (The notion ``cold collapse'' points 
to the interpretation of this approximation as the zero temperature 
limit, in which the classical pressure $p$ would vanish formally.) This allows 
one to study the nonlinear stage of a capillary--driven collapse under 
further simplifying assumptions: considering a radially symmetric particle 
distribution $\varrho(r, t)$ and taking $\lambda\to\infty$ (which can be 
phrased as the \emph{Newtonian limit}), one can solve exactly the 
equation for the radius $R(t)$ of a ring of particles. For our present 
purposes, we consider the particular configuration of an initial 
so-called top--hat density profile, 
\begin{equation} 
  \varrho(r, 0) = \left\{ 
    \begin{array}[c]{cc} 
      \varrho_\mathrm{h} , & r < L , \\ 
      0,  & r > L , 
    \end{array}\right. 
\end{equation} 
with initial size $L \ll \lambda$ so that the Newtonian limit 
holds at all times. The trajectory $R(t)$ of any single ring  
of matter can be obtained by specializing the general solution 
given in Ref.~\cite{Dominguez:2010} to this particular configuration: 
\begin{equation} 
  \label{eq:radius} 
  R(t) = R(0) \sqrt{ 1 - \frac{t}{\mathcal{T}} } , 
\end{equation} 
expressed in terms of Jeans' time as defined in 
Eq.~(\ref{eq:jeansT}). This result predicts that the whole matter collapses at 
the center simultaneously at a time $t=\mathcal{T}$. The ensuing divergence of 
this density field is actually preempted by the breakdown of the cold--collapse 
approximation at sufficiently high densities, because the force 
$\bF_\mathrm{short}$ halts further collapse.  
 
This approximation has also been used in Ref.~\cite{Dominguez:2010} in order 
to to predict 
qualitatively the time of collapse of spontaneous fluctuations around a 
homogeneous background density $\varrho_\mathrm{h}$: due to the underlying 
granular nature of the density field, in a surface area $\mathcal{A} (\ll 
\lambda^2)$  
there are unavoidable fluctuations with an amplitude $\approx 
\sqrt{\varrho_\mathrm{h} \mathcal{A}}$. Within the cold--collapse 
approximation, such  
initial fluctuations would collapse within the time   
\begin{equation} 
  \label{eq:Tcoll} 
  \mathcal{T}_\mathrm{coll} \approx  
  \mathcal{T} \ln \sqrt{\varrho_\mathrm{h} \mathcal{A}} . 
\end{equation} 
In view of the weak logarithmic dependence, this result predicts that for 
practical purposes fluctuations at all length scales collapse on 
a time scale $\sim \mathcal{T}$. 
 
In summary, the character of the dynamical evolution is set by two length 
scales: Jeans' length $K^{-1}$, above which the capillary attraction 
dominates over short--ranged repulsion, and the capillary length  
$\lambda$, above which the capillary attraction vanishes.  
Finally, the short--ranged repulsive forces set the minimum 
interparticle separation at which aggregation stops. 
 
%%%%%%%%%%%%%%%%%%%%%%%%%%%%%%%%%%%%%%%%%%%%%%%%%%%%%%%%%%%%%%%% 
%%%%%%%%%%%%%%%%%%%%%%%%%%%%%%%%%%%%%%%%%%%%%%%%%%%%%%%%%%%%%%%% 
 
\section{Brownian dynamics simulation} 
\label{sec:simulations} 
In this section we describe a theoretical approach for colloidal particles 
floating at an interface which is distinct from the one given by 
Eqs. (\ref{eq:short})-(\ref{eq:continuity}). To this end  
we consider a two--dimensional collection of $N$ disc--like particles enclosed 
in a square box of size  
$L\times L$ with periodic boundary conditions. Note that $L$ always denotes 
the system size which is either the side length of a square box or the 
diameter of a rotationally symmetric configuration.  
 
These particles interact via a short--ranged repulsive potential $V_{\rm 
  rep}(d)$ which avoids overlap. Specifically, we chose a shifted and cut--off 
Lennard--Jones potential mimicking the hard core repulsion between the 
particles: 
\begin{equation} 
  \label{eq:WCA} 
  \frac{V_\mathrm{rep}(d)}{k_\mathrm{B}T} 
  =\left\{ 
    \begin{array}[c]{cc} 
      \displaystyle  
      4\left[\left(\frac{2 R_0}{d}\right)^{12} 
      - \left(\frac{2 R_0}{d}\right)^{6} + \frac{1}{4}\right] , &  
      d \leq d_c \\ 
      & \\ 
      0 , & d \geq d_c, 
    \end{array}\right. 
\end{equation} 
where $R_0$ is the radius of the discs and $d_c/(2R_0)=2^{1/6} \approx 1.122$ 
the cut--off distance. In addition, a given particle is exposed to the 
capillary force from the other ones (see below). As noted before, we neglect 
any hydrodynamic interactions, a systematic treatment of which is beyond the 
scope of the present work. 
 
Due to the coupling to the two fluid phases, the discs experience stochastic 
forces and their motion is overdamped leading to dissipation. This kind of 
motion is described by Brownian dynamics. Applying the Ermak 
algorithm~\cite{Ermak:1975} and using the notation introduced in 
Ref.~\cite{Allen:1987}, the integration of the corresponding Langevin equation 
yields for the change of the position of each disc 
\begin{equation} 
  \label{eq3} 
  \dot{\br}= \Gamma \bar{\bforce} + \mathring{\br} , 
\end{equation} 
where $\mathring{\br}$ is a random velocity with 
$\langle\mathring{\br}\rangle=0$ and 
$\langle\mathring{\br}_\mathrm{i}(t) \mathring{\br}_\mathrm{j}(t^{\prime})\rangle= 
\delta(t-t^{\prime})\delta_\mathrm{ij}\dblone$~\cite{Allen:1987}.      
%%%%%%%%%%%%%%%%%%% 
The probability distribution of $\mathring{\br}$ is 
taken to be uniform with a width $w$ depending on the time step $\Delta t$: 
\begin{equation} 
  \label{eq:width} 
  w(\Delta t)=\sqrt{2D \Delta t}, 
  \qquad 
  D := k_\mathrm{B}T\Gamma . 
\end{equation} 
It turns out that choosing a Gaussian distribution instead 
does not change the results of the present study.  
The flow velocity is $\bv = \langle\dot{\br}\rangle$ and the 
force per particle is $\bar{\bforce}=(\bF_\mathrm{rep}+\bF_\mathrm{cap})/ 
\varrho$ (see Eq.~(\ref{eq:flow})). 
 
This particle--based model constitutes an alternative description of the 
system discussed in the previous section. The random displacements of the 
colloids in combination with the action of the short--ranged repulsive 
forces give rise to the macroscopic colloidal pressure $p$, as used in 
Eq. (\ref{eq:short}), with the corresponding equation of state 
$p=p(\varrho)$ which, in principle, could be determined within this model.  
 
Generically, the  
capillary force $\bF_\mathrm{cap}$ on a single particle involves a very 
large number of neighboring particles due to its  
long range $\lambda\gg R_0$. Furthermore, we are also interested in the case 
that $\lambda$ exceeds the side length $L$ of the system. Therefore, the 
capillary interactions are calculated  
using the so--called particle--mesh algorithm (PM), as applied in cosmological 
simulations which face a similar difficulty (see, e.g., 
Refs.~\cite{Hockney:1988,Deserno:1998,Knebe}). The basic idea is to solve 
Eq.~(\ref{eq:YL}) via Fourier transformation ($\mathcal{F}[g(\br)]\equiv \int 
d^2r\, e^{-i\br\cdot\bk}g(\br)=\hat{g}(\bk)$) on a mesh (or grid) which 
discretizes the plane. Accordingly, Eq.~(\ref{eq:YL}) leads to  
\addtocounter{equation}{+1} 
\begin{align*}{\tag{\theequation a}} 
  \label{eq4} 
  \left(-k^2 -\frac{1}{\lambda^2}\right)\widehat{U}(\bk) &= 
  -\frac{f}{\gamma}\widehat{\varrho}(\bk) 
\end{align*} 
so that 
\begin{equation*}{\tag{\theequation b}} 
  U(\br) = \mathcal{F}^{-1}\left[-\frac{f}{\gamma}\,\widehat{G}(\bk)\, 
    \widehat{\varrho}(\bk)\right] 
\end{equation*} 
where $\mathcal{F}$ and $\mathcal{F}^{-1}$ denote the Fourier transformation 
and its inverse, respectively. $\widehat{G}(\bk)=-1/(k^2 
+\lambda^{-2})$ is Green's function for this differential equation. 
The Fourier transformation can be easily implemented using standard Fast 
Fourier Transform (FFT) routines.  
The Fourier--transformed density field $\widehat{\varrho}(\bk)$ is computed 
from a given $\varrho(\br)$ by constructing a discretized version of the 
particle density distribution. Also $\widehat{G}(\bk)$ is discretized such 
that it respects the periodic boundary conditions. 
 
In order to discretize the density, we have employed so--called mass assignment 
schemes using two different versions: the so--called Cloud in Cell (CIC) 
method, in which each particle is distributed between four neighboring 
cells, and the so--called Triangular Shaped Cloud (TSC) method using nine 
neighboring cells. The latter one yields an an enhanced smoothening  
of the particle distribution. Depending on the mesh size of the discretization 
grid, either of these standard methods may be the more accurate choice.  
For a grid with a mesh size (somewhat) smaller than the particle size, the 
distribution to four (CIC) or nine (TSC) neighboring cells corresponds to 
a smeared patch the size of which is approximately the actual particle size. 
For the discretized Green's function we use 
\begin{equation} 
  G(k_\mathrm{lm})= \frac{1}{-\sin^2(k_\mathrm{x}/2)- 
    \sin^2(k_\mathrm{y}/2)-\lambda^{-2}}  
\end{equation} 
with  
\begin{equation} 
  k_\mathrm{x}=\frac{2\pi l}{L}, \quad k_\mathrm{y}=\frac{2\pi m}{L},  
  \qquad l,m \in \mathbb{Z}. 
\end{equation} 
The capillary force is obtained from the interfacial deformation field 
via Eq.~(\ref{eq:Fcap}): 
\begin{align} 
  \label{eq:Fgrid} 
  \frac{\bF_\mathrm{cap}(\br)}{\varrho(\br)} &= \nabla [f U (\br)]\nonumber\\ 
  &=\mathcal{F}^{-1}\left[\frac{f^2}{\gamma}\,i\bk\, 
    \widehat{G}(\bk)\,\widehat{\varrho}(\bk)\right] . 
\end{align} 
This yields the discretized capillary force \emph{per particle}  at 
the grid points. In order to obtain the actual force at the particle position, 
one has to interpolate the discretized force using the inverse of the chosen 
mass assignment scheme.  
 
This method offers two important advantages. First, it is fast because the 
number of operations of the PM algorithm scales like $N_\mathrm{g} \log 
N_\mathrm{g}$, where $N_\mathrm{g}$ is the number of nodes of the 
grid. Second, this method guarantees an appropriate incorporation of the 
long--ranged potential, i.e., without introducing any cutoffs together with a 
high accuracy down to small distances of the order of the grid spacing 
$r_\mathrm{c}$. We have chosen $r_\mathrm{c}=R_0$.      
 
In order to save computation time by reducing the number of integration steps, 
and, even more important, in order to avoid large displacements $\Delta\br$ by 
the random velocity process, the time step is not fixed during the simulation, 
but it is calculated in each iteration step from the requirement that the 
maximum displacement due to the force acting on any individual colloid should 
not exceed $R_0/4$. This corresponding maximum time step is then used in order 
to calculate  
the width of the distribution for the random displacement according to 
Eq.~(\ref{eq:width}). We note that this displacement, due to the thermal 
fluctuations, can be larger than the nominal value of $R_0/4$. This cut-off  
distance appears to be small for simulation boxes with $L\sim 500 R_0$. This is, 
however, a compromise which on one hand ensures reasonable computational 
costs, and on the other hand avoids large values of repelling forces.  
%%%%%%%%%%%%% 
Such forces emerge if a particle is displaced randomly on top of another 
particle and encounters the hard core of the repulsive potential. At such small 
distances, the $r^{-12}$ term in the  
potential would generate large forces which in turn would lead to large 
velocities, yielding the approximation of an overdamped motion and thus the 
Langevin equation for the position inapplicable.   
%%%%%%%%%%%%% 
 
The structure of the simulation algorithm can be summarized by the following 
main steps:     
\begin{enumerate} 
\item random placement of particles on a plane without overlap; 
\item assignment of particles to a grid 
  by means of a mass 
  assignment scheme, resulting in a number density field defined on the grid;  
\item Fourier transformation of this density field by application of 
  standard FFT; 
\item computation of the capillary force field on the grid according to 
  Eq.~(\ref{eq:Fgrid});  
\item interpolation of this force field to the positions of the particles using 
  the inverse mass assignment scheme;  
\item calculation of short--ranged repulsive forces by summation over 
  all particles within one box size (minimal image convention); 
\item calculation of the maximum possible time step; 
\item integration of the equations of motion by using Ermak's algorithm for 
  Eq.~(\ref{eq3}) and loop back to step 2. 
\end{enumerate} 
%%%%%%%%%%% 
This algorithm approximates Eq.~(\ref{eq:continuity}) in terms of individual 
particle positions which, when assigned to a density grid, represent the 
approximate evolution of $\varrho(\br,t)$. 
%%%%%%%%%%%   
 
%%%%%%%%%%%%%%%%%%%%%%%%%%%%%%%%%%%%%%%%%%%%%%%%%%%%%%%%%%%%%%%% 
%%%%%%%%%%%%%%%%%%%%%%%%%%%%%%%%%%%%%%%%%%%%%%%%%%%%%%%%%%%%%%%% 

\section{Results} 
\label{sec:results} 
 
In order to probe the predictions of the theoretical model 
reviewed in Sec.~\ref{sec:theory}, for two types of initial configurations 
with periodic boundary conditions we have carried out the simulations 
described above for discs of radius $R_0$ with a short--ranged repulsion 
according to Eq.~(\ref{eq:WCA}):  
\begin{itemize} 
\item a random homogeneous distribution of non-overlapping discs 
  in a square box of size $L\times L$, 
\item a rotationally invariant configuration forming a disc of radius $L$ 
  surrounded by vacuum, i.e., embedded in an empty square box of size 
  $L^{\prime} > 2 L$.  
\end{itemize} 
We have chosen the temperature to be low enough so that these configurations are 
unstable, i.e., $\lambda K > 1$ (see Eq.~(\ref{eq:Kc})), and the particles 
therefore tend to lump together under the driving capillary force. 
The effects specifically due to the  
long--ranged nature of this interaction can be studied by comparing 
simulation data for various values of $\lambda$. By changing the screening 
parameter $\lambda$ the interparticle attraction can be tuned from 
being de facto short--ranged ($\lambda$ of the order of a few colloid 
diameters, such that the minimum image convention applies) to being 
long--ranged ($\lambda \to \infty$, for which the attraction is 
formally analogous to 2D Newtonian gravity). At sufficiently low temperatures 
$T$, the behavior of the system is determined by minimizing its potential 
energy. Thus, for low $T$ the final state for any system of size $L$ is a 
single cluster with maximum close packing of the particles determined by the 
interparticle repulsion (e.g., due to the hard core of the discs). This 
cluster will contain most particles. However, the range of the force 
determines how and on which time scale this final state is reached. 
\begin{itemize} 
\item $R_0, L \ll \lambda$; in this limiting case it is expected that 
  phenomena occur which are similar to the gravitational collapse studied in 
  the astrophysical context. In the so--called Newtonian 
  limit ($\lambda\to\infty$) the theoretical analysis summarized  in 
  Sec.~\ref{sec:theory} suggests that Jeans' time $\mathcal{T}$ (see 
  Eq.~(\ref{eq:jeansT})) sets the corresponding characteristic time scale of 
  evolution. 
\item $R_0 \lesssim \lambda \ll L$; in this limit the behavior of the system 
  is expected to be similar to that of a common fluid and the 
  corresponding mean--field description would be analogous to a van--der--Waals 
  type equation of state. The temporal evolution is expected to follow the 
  scenario of spinodal decomposition because the initial density $\varrho_h$ 
  lies in the unstable region between the liquid and vapor coexistence 
  densities.    
\end{itemize} 
One of the goals of the present simulations is to study the crossover between 
these two limiting behaviors.  
 
We have performed several simulations based on the set of parameter values 
shown in Table~\ref{tab1}, which all correspond to an  
experimentally realistic choice of parameters. Unless stated otherwise, all 
simulations are based on the parameter values quoted in Table \ref{tab1}.  
We always employ initial conditions with such diluted 
densities that the equation of state of an ideal gas can be applied in order 
to obtain a rough estimate of Jeans' length from Eqs.~(\ref{eq2}) and 
(\ref{eq:jeansK}): 
\begin{equation} 
  \label{eq:k2} 
  K^2 \approx \frac{f^2 \varrho_h}{\gamma k_\mathrm{B} T} 
  = 2\pi \varrho_h \frac{V_0}{k_\mathrm{B} T} . 
\end{equation} 
\begin{table}[ht] 
  \begin{tabular}{|c|c|c|} 
    \hline 
     number of colloids&$N$&$2500$\\ 
     \hline 
     colloid radius&$R_0$&$10\, \mu$m\\ 
     \hline 
     system size&$L$&$5\,$mm\\ 
     \hline 
     capillary length&$\lambda$&$2.7\,$mm\\ 
     \hline 
     strength of capillary force&$V_0$&$0.89\,k_\mathrm{B} T$\\ 
     \hline 
     particle mobility&$\Gamma$&$10.6\,\times 10^{-6}$s/kg\\ 
     \hline 
     mean particle separation&$L/N^{1/2}$&$100\, \mu$m\\ 
     \hline 
     Jeans' length&$1/K$&$43.8\, \mu$m\\ 
     \hline 
     Jeans' time&$\mathcal{T}$&$40780\,$s\\ 
     \hline 
  \end{tabular} 
  \caption{\label{tab1} Default values of the parameters used for all  
    simulations presented here and certain quantities deduced from them. The 
    value of $\lambda$ corresponds to the capillary length of a pure water--air 
    interface and the value of $\Gamma$ to the effective mobility of a sphere 
    half immersed in water at room temperature~\cite{Dominguez:2010}.}  
\end{table} 
 
\subsection{Random homogeneous distribution as initial configuration} 
 
\begin{figure}[ht] 
  \epsfig{file=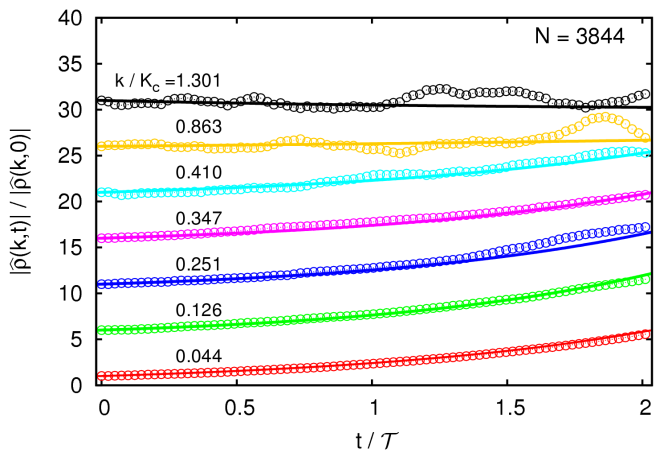,height=\linewidth,angle=270} 
  \caption{\label{fig_2}Evolution of %unstable 
    Fourier modes 
    $\widehat{\varrho}(|\bk|<K_\mathrm{c},t=0)$ of the initial density 
    distribution (averaged over shells of constant $k=|{\bf k}|$ and each 
    shifted upwards by $5$ for clarity) in a simulation for  
    a $L\times L$ box, $L=620 R_0$, with $N=3844$ particles. According to 
    Eq.~(\ref{eq:Kc}) and Table \ref{tab1}, $K_\mathrm{c}=22.8\, 
    \mathrm{mm}^{-1}$ and  
    $\mathcal{T}=40780 s$. The error bars concerning the statistical errors, 
    as obtained from averaging within small time intervals of the order 
    of $\Delta t \approx \mathcal{T}/40$ and over $20$ runs of initial 
    configurations, are smaller than the symbol size of the simulation 
    data. The full lines provide the corresponding theoretical prediction 
    $\exp(t/\tau(k))$ as  given by Eqs.~(\ref{eq:linearEvol}) and 
    (\ref{eq:tau}). The mean initial number density is 
    $\varrho_h=10^{-2}R_0^{-2}$.}   
\end{figure}

First  we address the quantitative prediction in Eqs.~(\ref{eq:linearEvol}) 
and (\ref{eq:tau}) 
concerning the early evolution of the initial fluctuations. In 
Fig.~\ref{fig_2} we show the growth of the amplitudes of %unstable 
various modes of the Fourier--transformed initial density
field, normalized by their values at  $t=0$. The evolution is expected to
follow the exponential dependence given in  Eqs.~(\ref{eq:linearEvol}) and
(\ref{eq:tau}). 
According to these equations, unstable modes are characterized by the
condition $k < K_\mathrm{c} $ 
(see Eq.~(\ref{eq:Kc})) with amplitudes increasing with time. We identify
the fastest growing mode as the one with $k = 0.126\, K_{\mathrm c}$. 
Modes with $k > K_{\mathrm c}$, the amplitudes of which decrease with time, are
stable.
Up to $t \sim 1.5\mathcal{T}$,
the agreement with this theoretical prediction is very good. Thus the
linearized stability analysis is reliable up to the characteristic time
$\mathcal{T}$ of the system.  
 
\begin{figure}[ht] 
  \begin{minipage}{0.32\linewidth} 
    \begin{center} 
      $t=\mathcal{T}$  
    \end{center} 
  \end{minipage} 
  \begin{minipage}{0.32\linewidth} 
    \begin{center} 
      $t=3\mathcal{T}$  
    \end{center} 
  \end{minipage} 
  \begin{minipage}{0.32\linewidth} 
    \begin{center} 
      $t=5\mathcal{T}$  
    \end{center} 
  \end{minipage} 
  \begin{minipage}{\linewidth} 
    \vspace{0.5cm} 
  \end{minipage} 
  \begin{minipage}{0.02\linewidth} 
    \begin{sideways} 
      $\lambda/L=0.02$ 
    \end{sideways} 
  \end{minipage} 
  \begin{minipage}{0.968\linewidth} 
    \epsfig{file=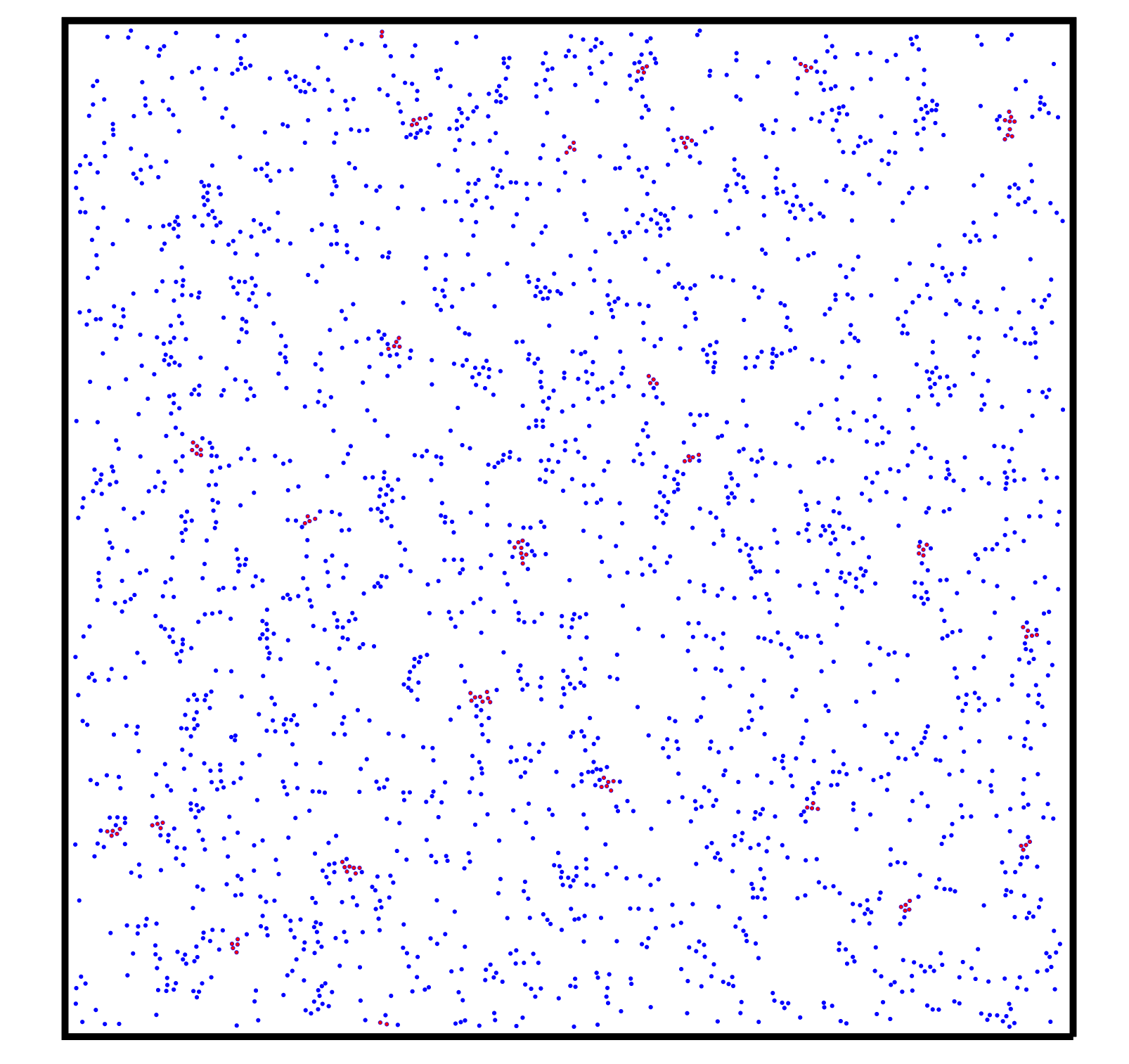,width=.32\linewidth} 
    \epsfig{file=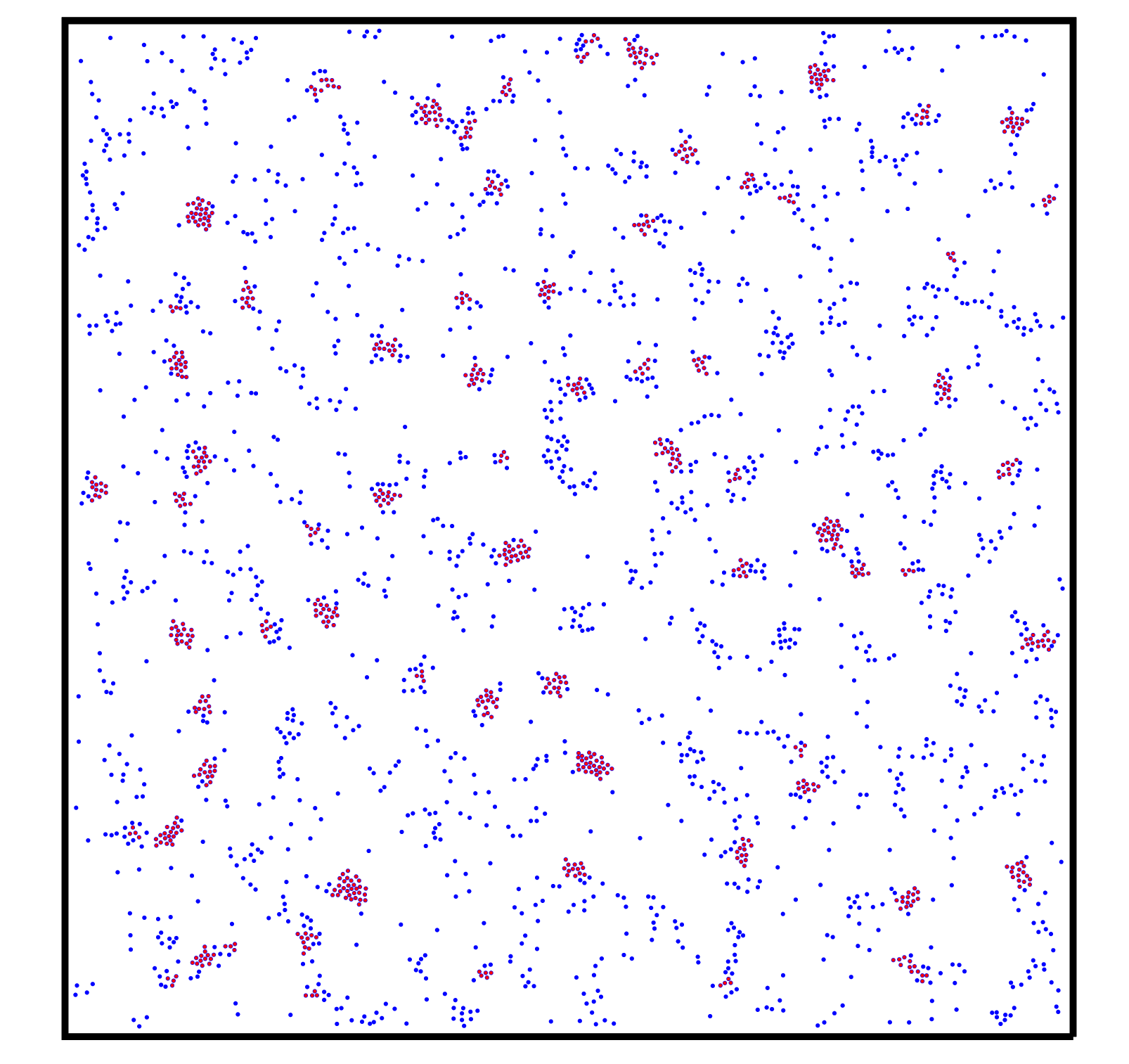,width=.32\linewidth} 
    \epsfig{file=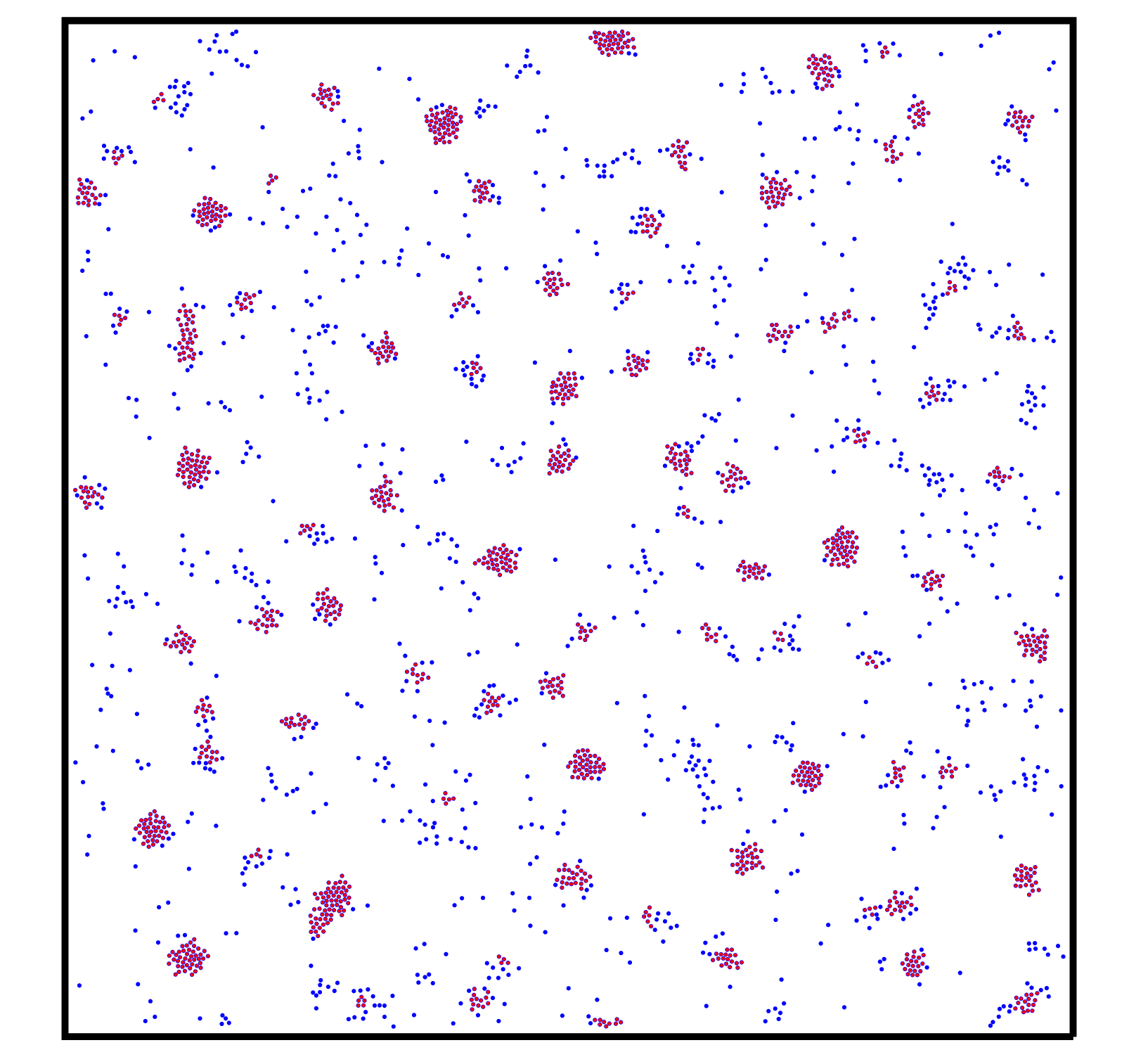,width=.32\linewidth} 
  \end{minipage} 
  \begin{minipage}{0.02\linewidth} 
    \begin{sideways} 
      $\lambda/L=0.05$ 
    \end{sideways} 
  \end{minipage} 
  \begin{minipage}{0.968\linewidth} 
    \epsfig{file=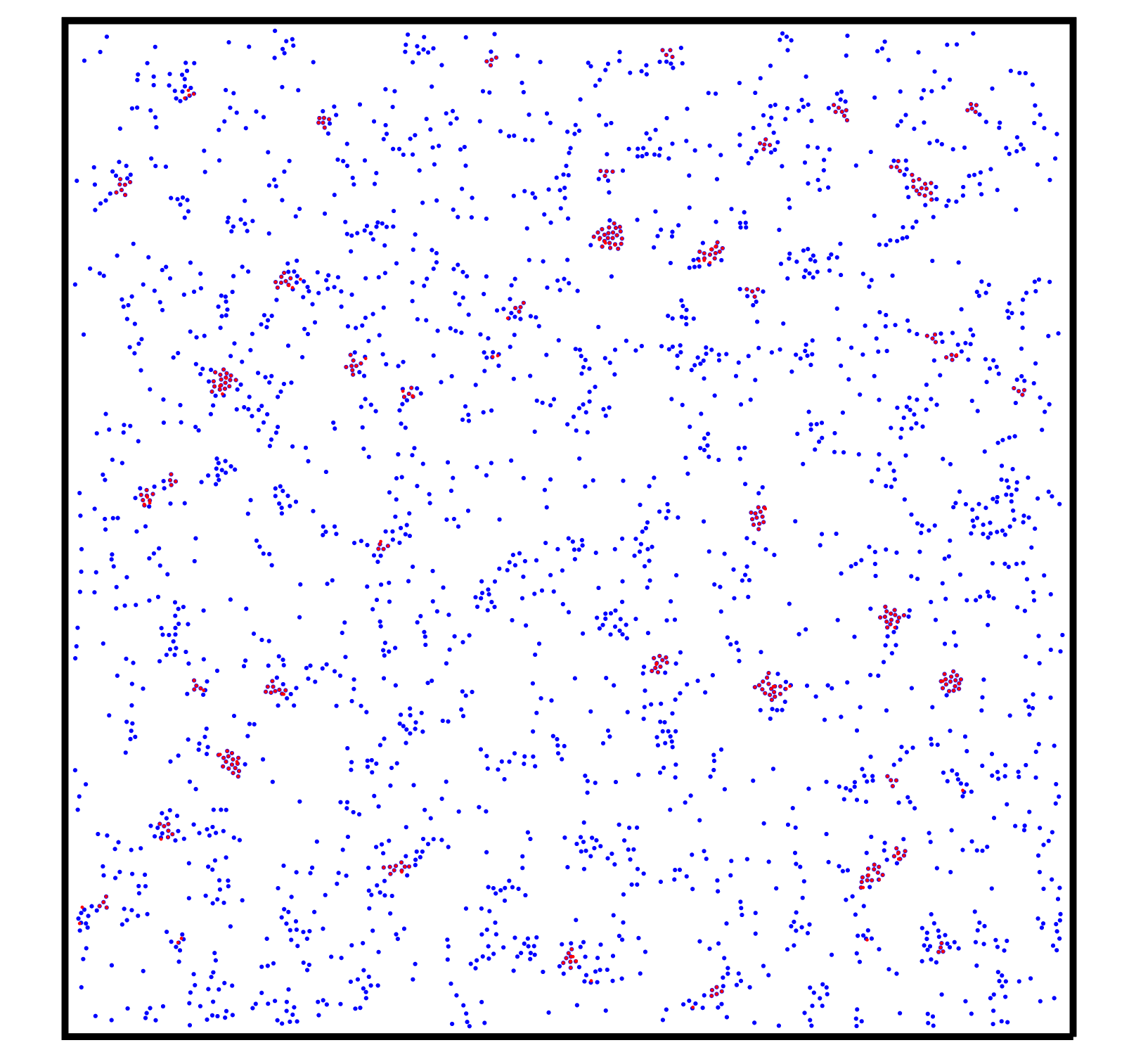,width=.32\linewidth} 
    \epsfig{file=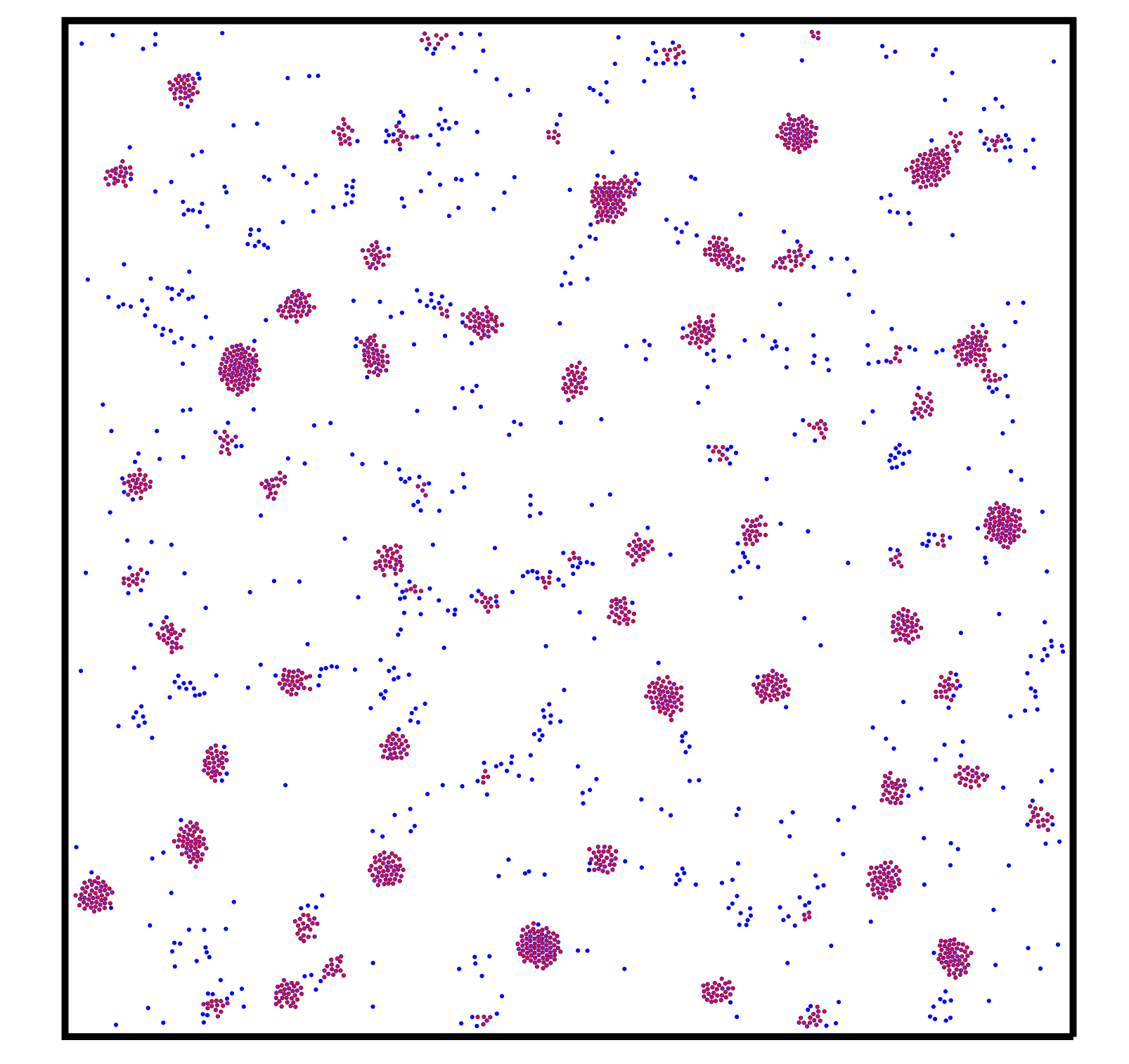,width=.32\linewidth} 
    \epsfig{file=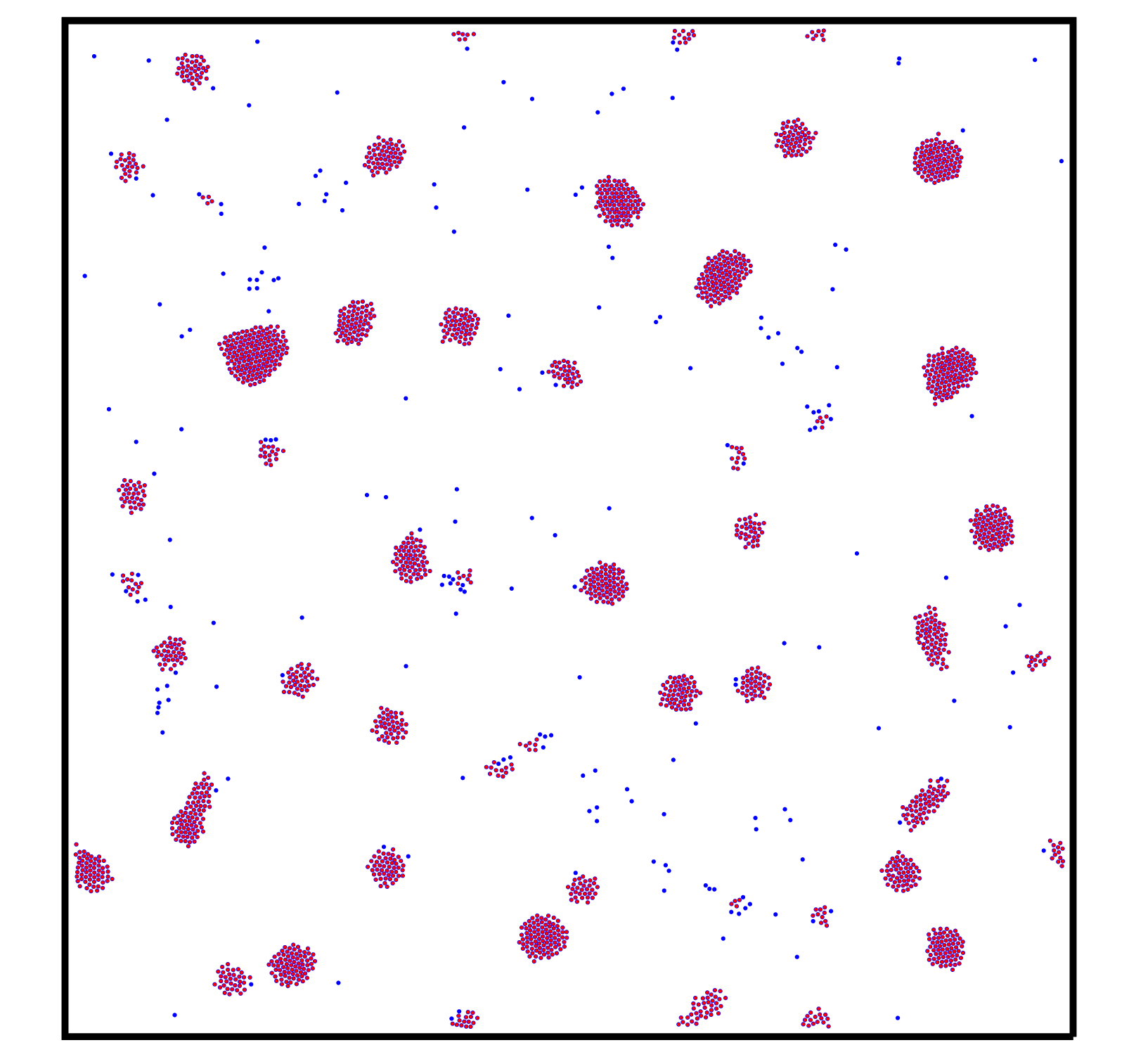,width=.32\linewidth} 
  \end{minipage} 
  \begin{minipage}{0.02\linewidth} 
    \begin{sideways} 
      $\lambda/L=0.1$ 
    \end{sideways} 
  \end{minipage} 
  \begin{minipage}{0.968\linewidth} 
    \epsfig{file=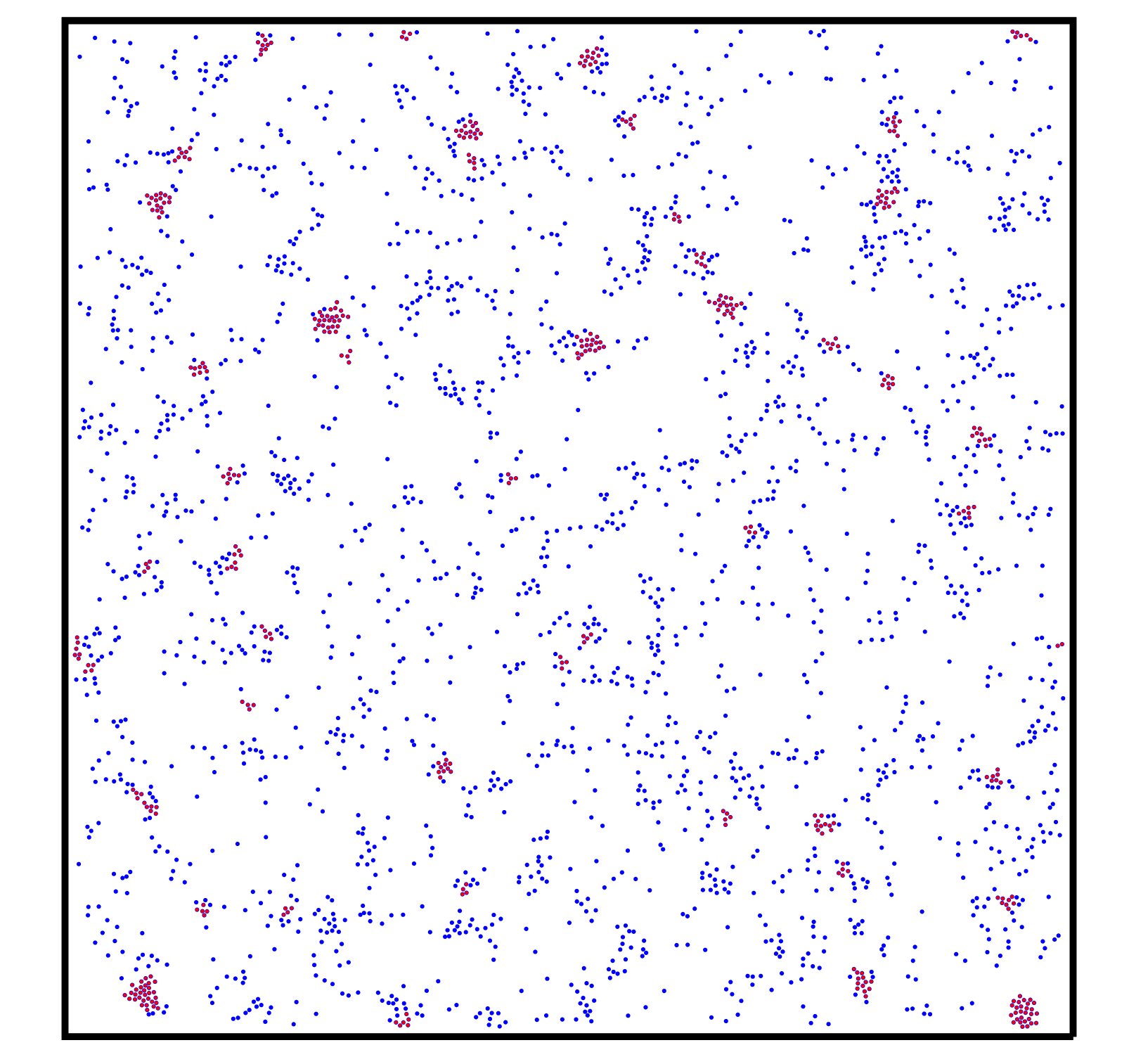,width=.32\linewidth} 
    \epsfig{file=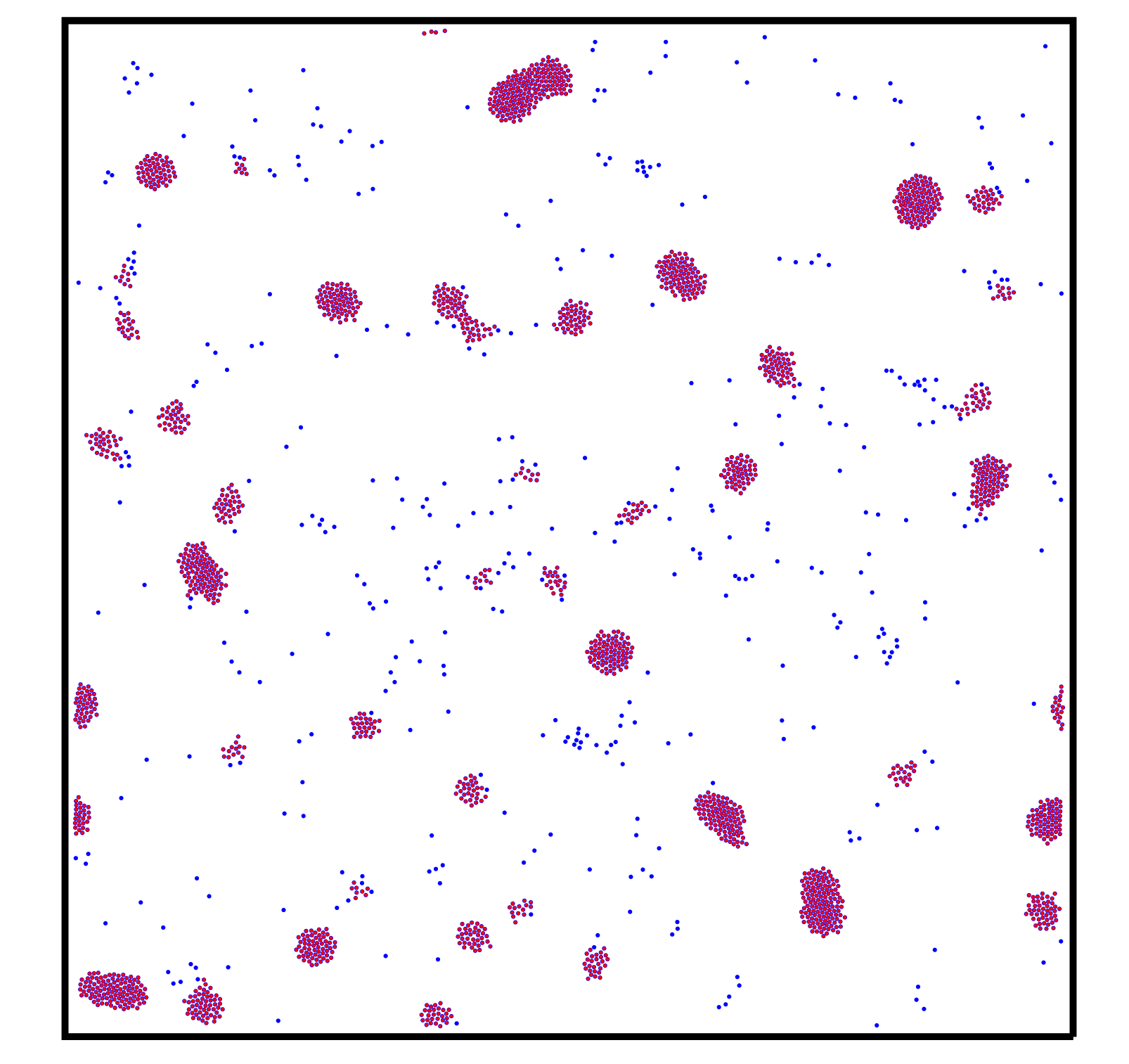,width=.32\linewidth} 
    \epsfig{file=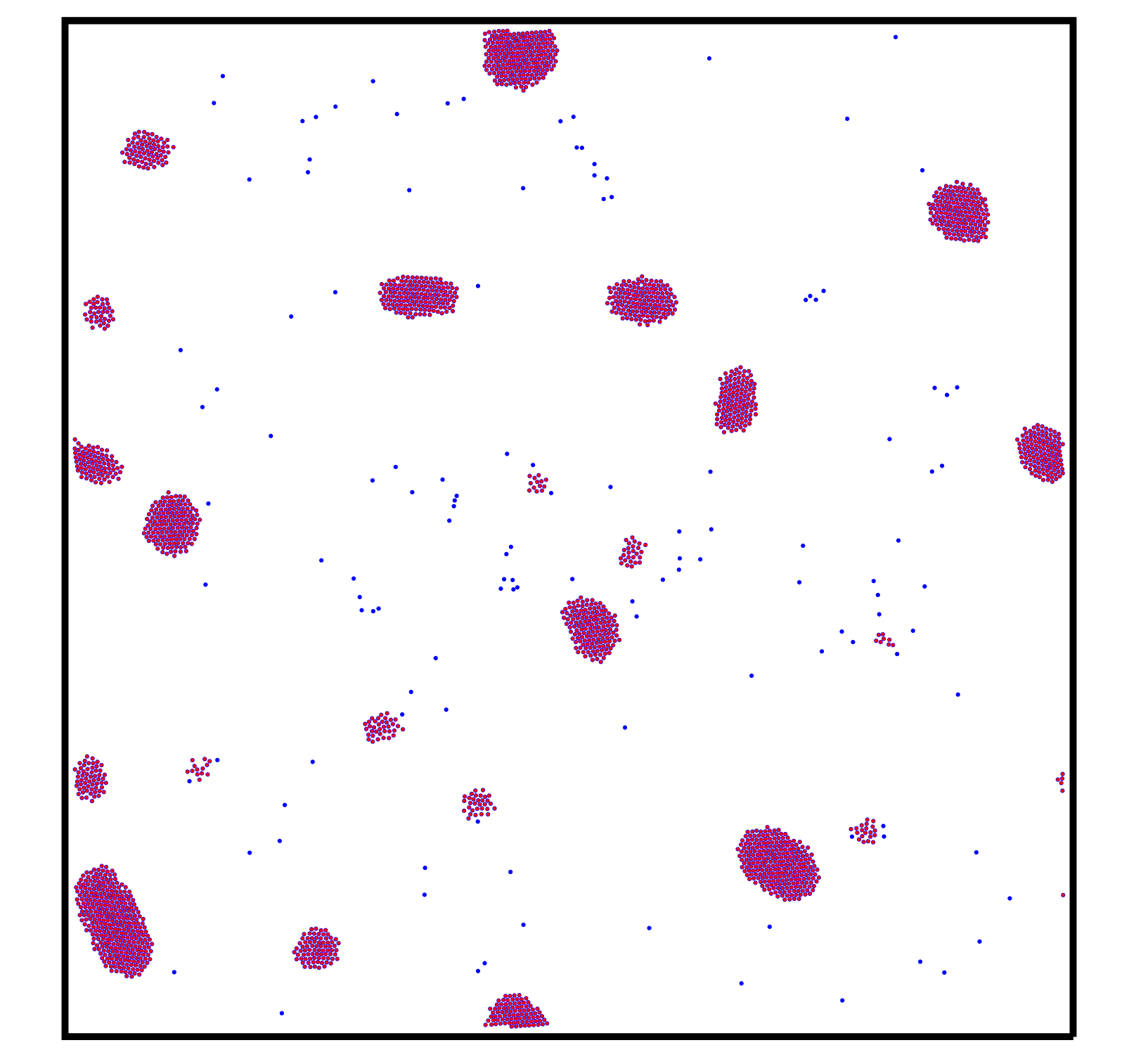,width=.32\linewidth} 
  \end{minipage} 
  \caption{\label{fig_5}Snapshots of the particle distribution in 
    $L\times L$ boxes with $L=500 R_0$ at times $\mathcal{T}$, $3 
    \mathcal{T}$, and $5 \mathcal{T}$ for a ``short--ranged'' 
    capillary attraction with $\lambda/L=0.02$, $\lambda/L=0.05$, and 
    $\lambda/L=0.1$ (from top to bottom). The number of particles is 
    $N=2500$. Particles belonging to a cluster are colored in red.} 
\end{figure} 
 
\begin{figure}[ht] 
  \begin{minipage}{0.32\linewidth} 
    \begin{center} 
      $t=\mathcal{T}$  
    \end{center} 
  \end{minipage} 
  \begin{minipage}{0.32\linewidth} 
    \begin{center} 
      $t=3\mathcal{T}$  
    \end{center} 
  \end{minipage} 
  \begin{minipage}{0.32\linewidth} 
    \begin{center} 
      $t=5\mathcal{T}$  
    \end{center} 
  \end{minipage} 
  \begin{minipage}{\linewidth} 
    \vspace{0.5cm} 
  \end{minipage} 
  \begin{minipage}{0.02\linewidth} 
    \begin{sideways} 
      $\lambda/L=0.54$
    \end{sideways} 
  \end{minipage} 
  \begin{minipage}{0.968\linewidth} 
    \epsfig{file=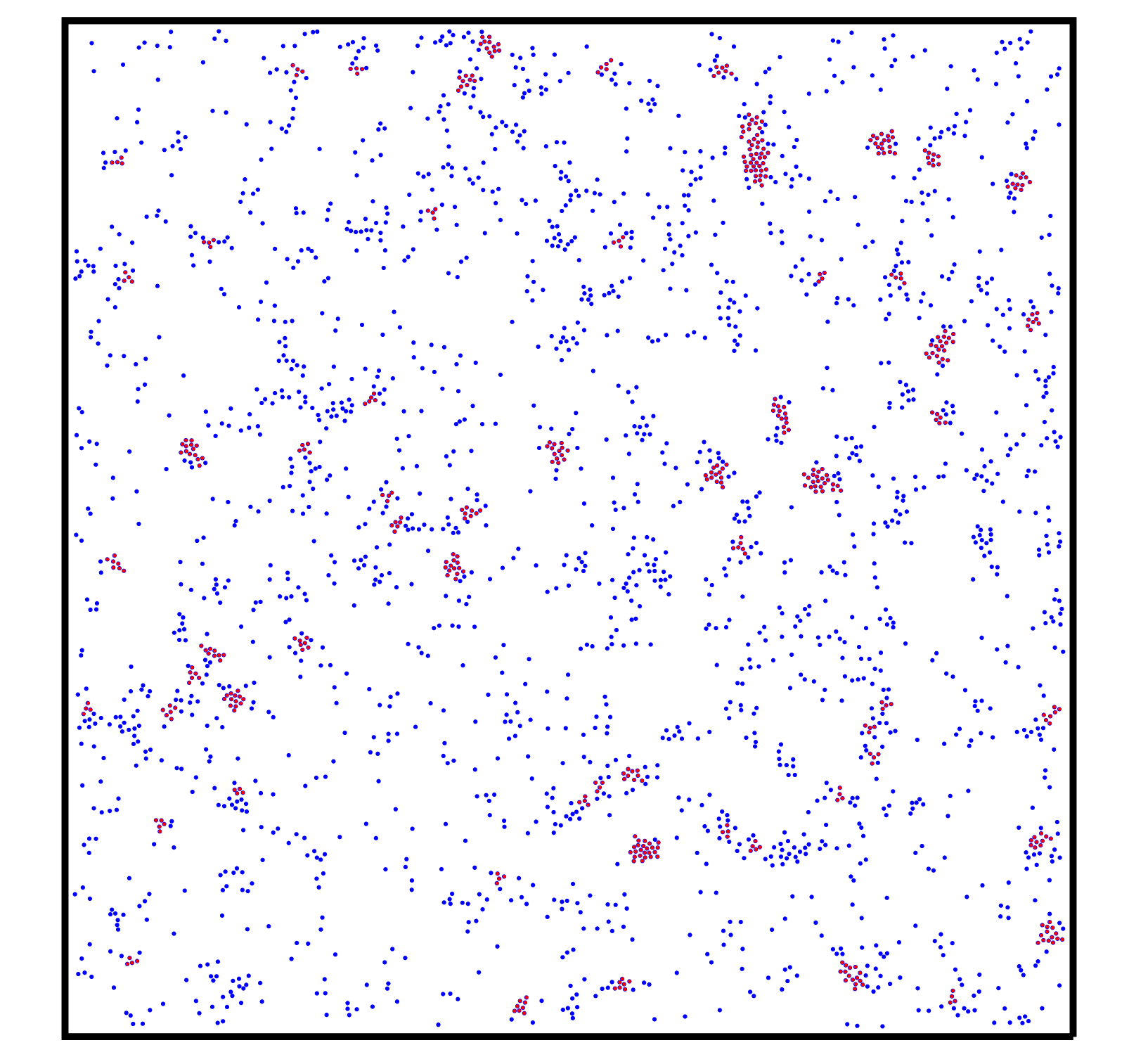,width=.32\linewidth} 
    \epsfig{file=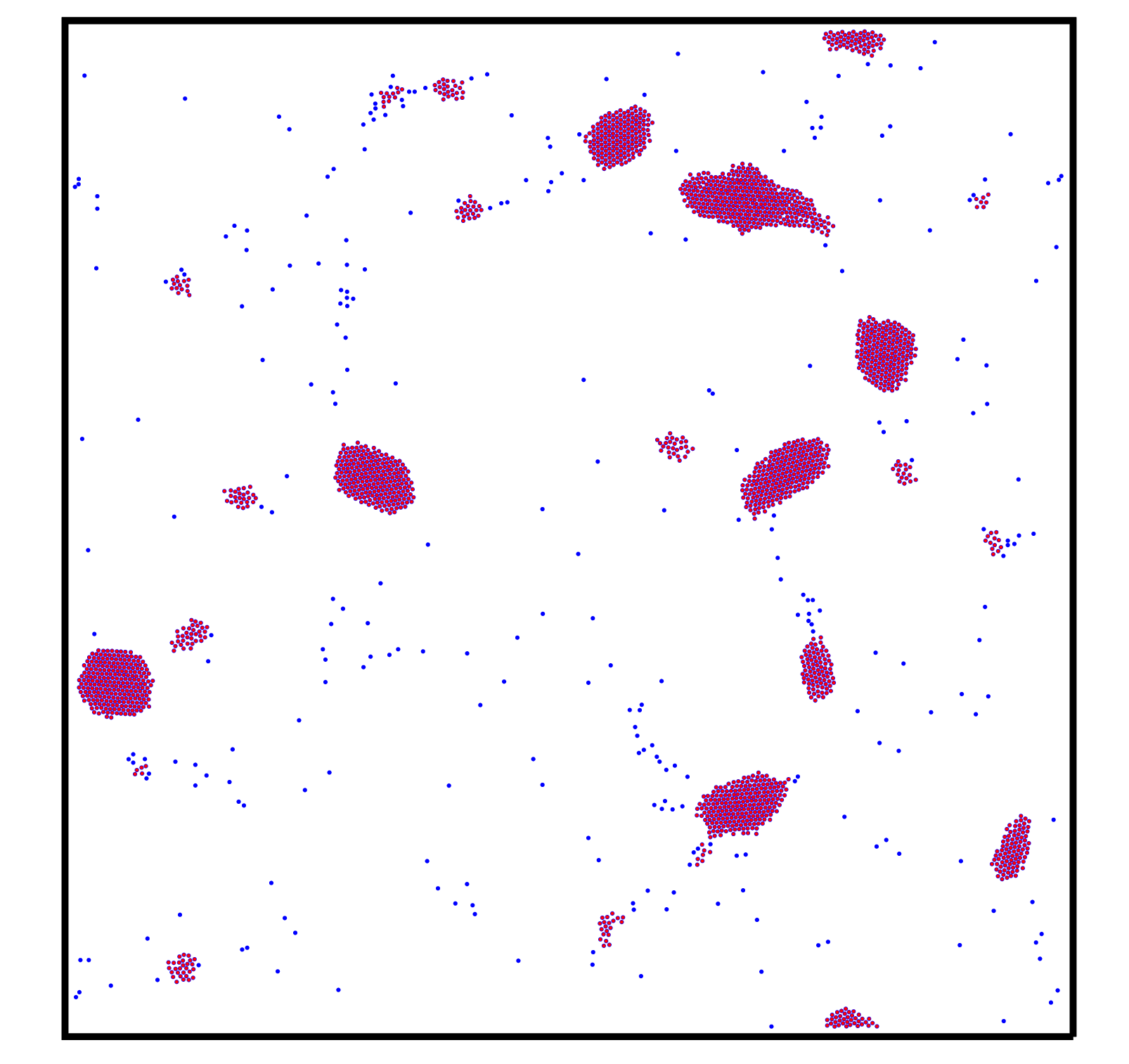,width=.32\linewidth} 
    \epsfig{file=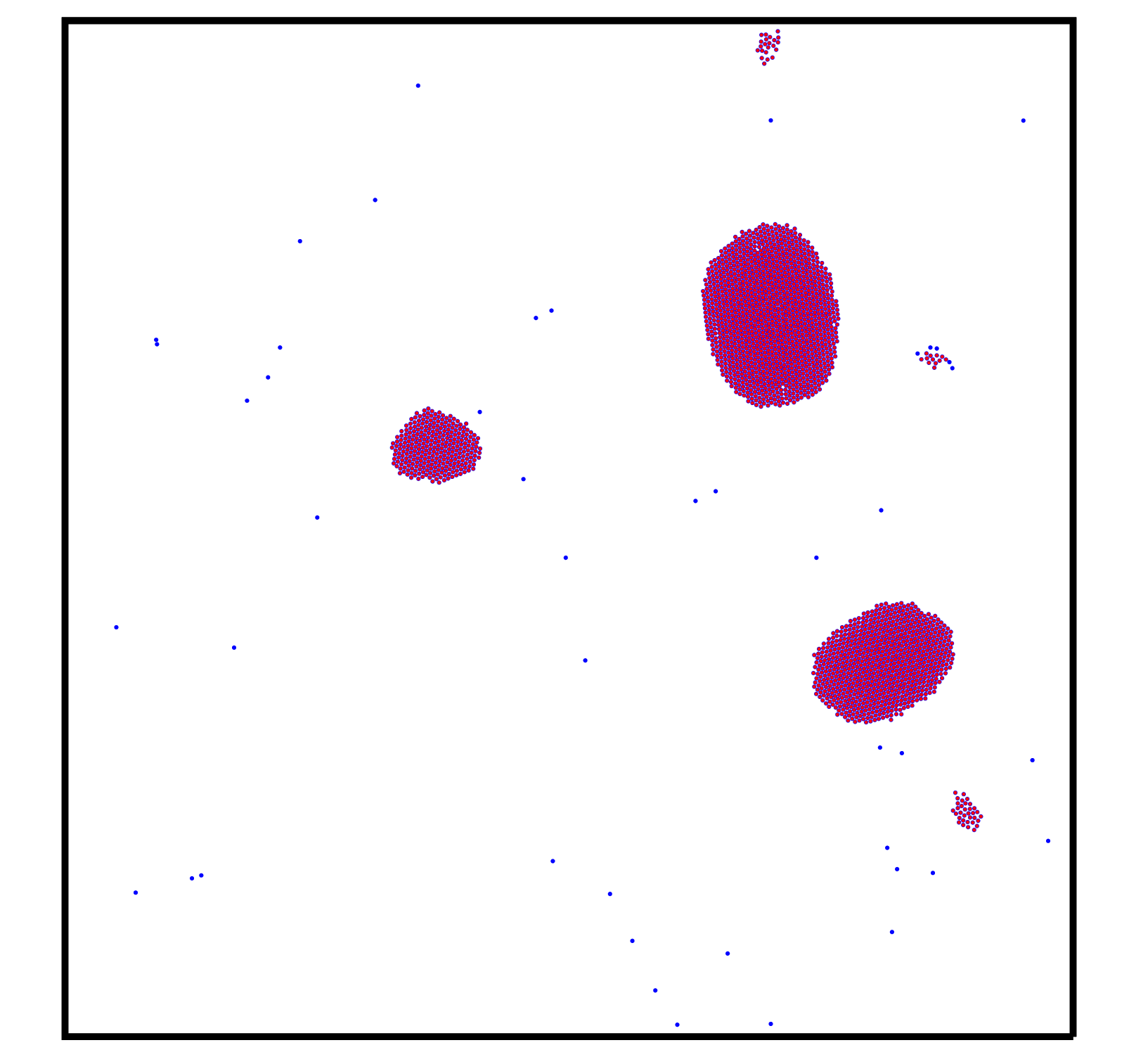,width=.32\linewidth} 
  \end{minipage} 
  \begin{minipage}{0.02\linewidth} 
    \begin{sideways} 
      $\lambda/L=3.0$ 
    \end{sideways} 
  \end{minipage} 
  \begin{minipage}{0.968\linewidth} 
    \epsfig{file=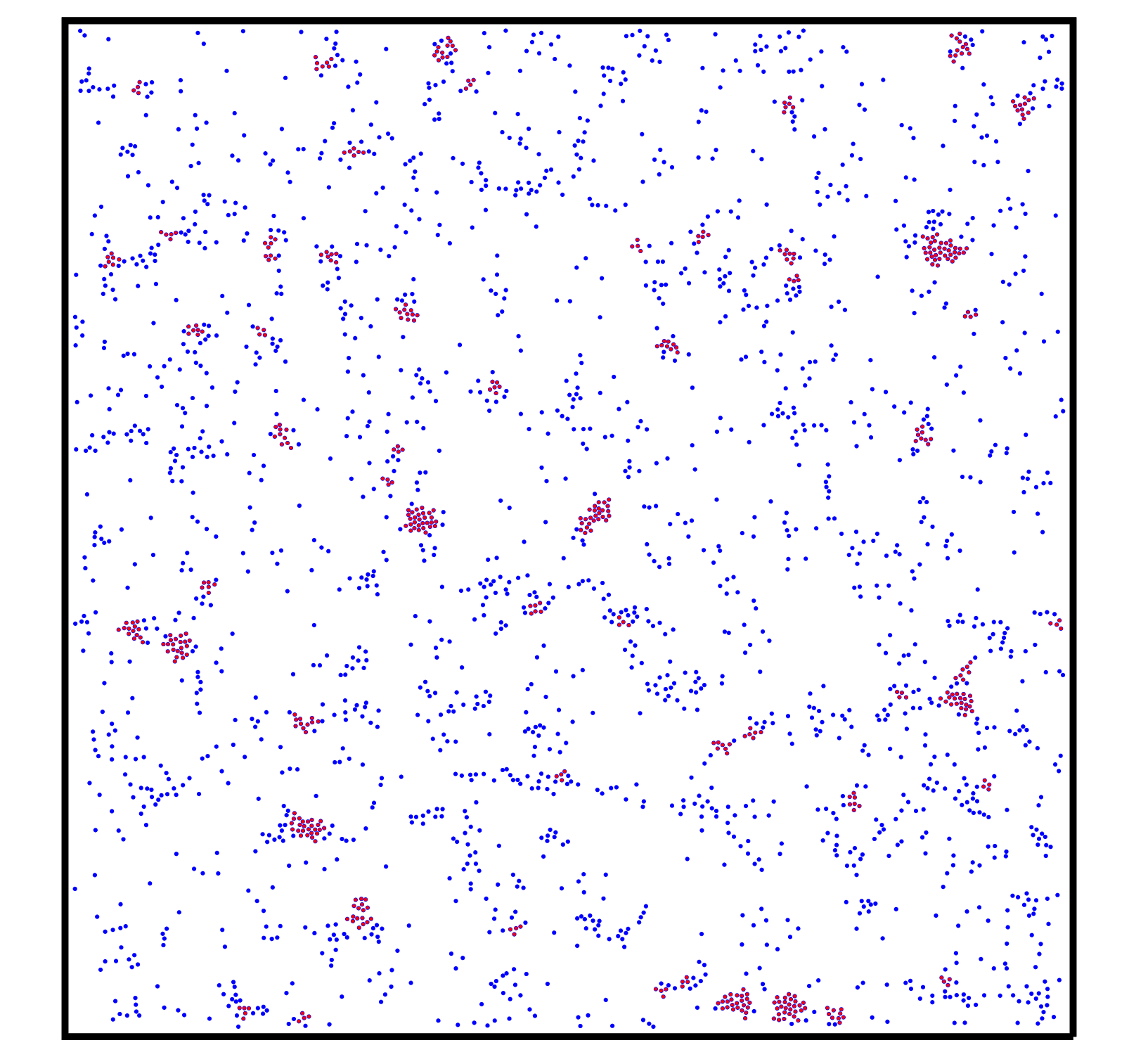,width=.32\linewidth} 
    \epsfig{file=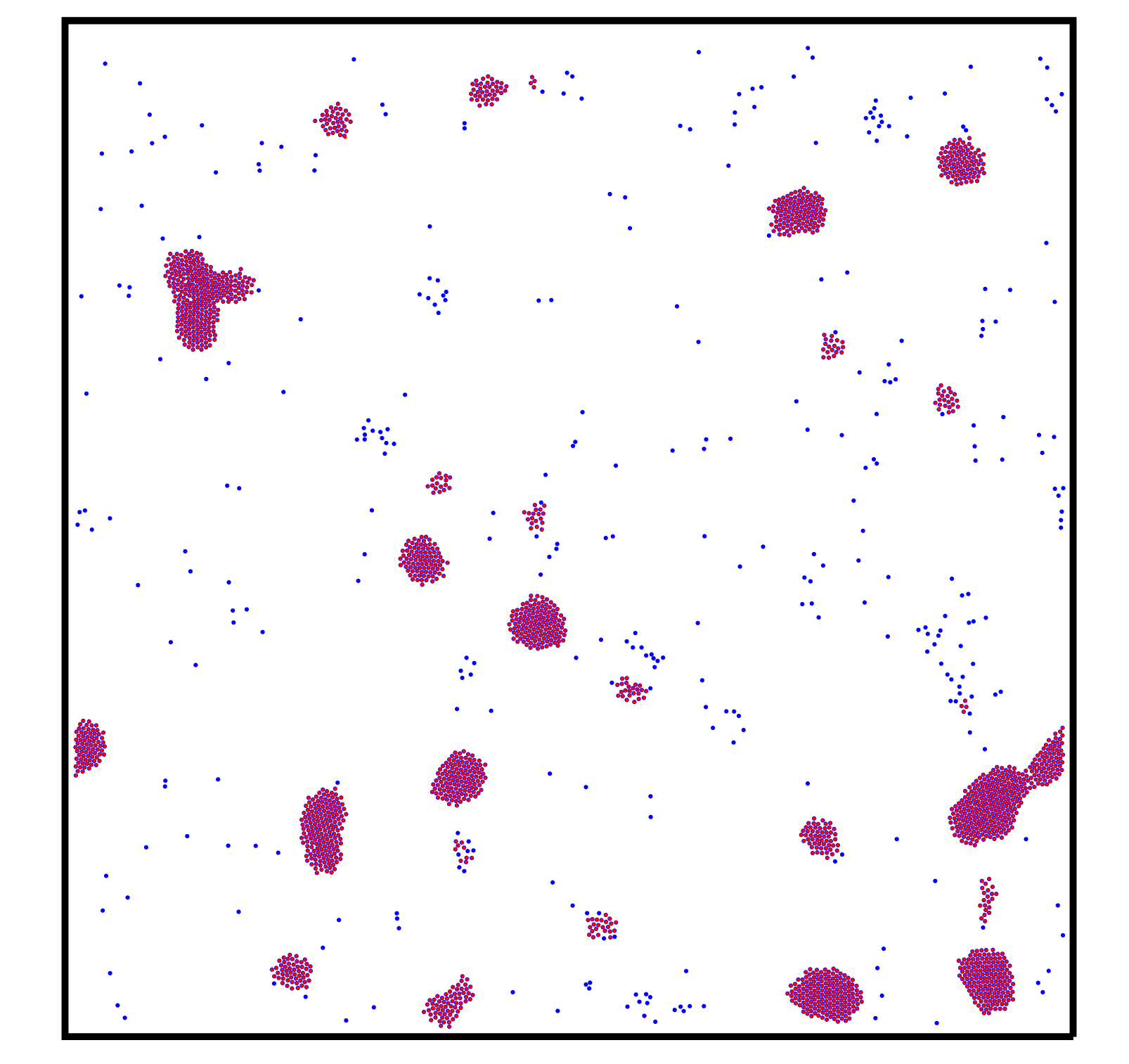,width=.32\linewidth} 
    \epsfig{file=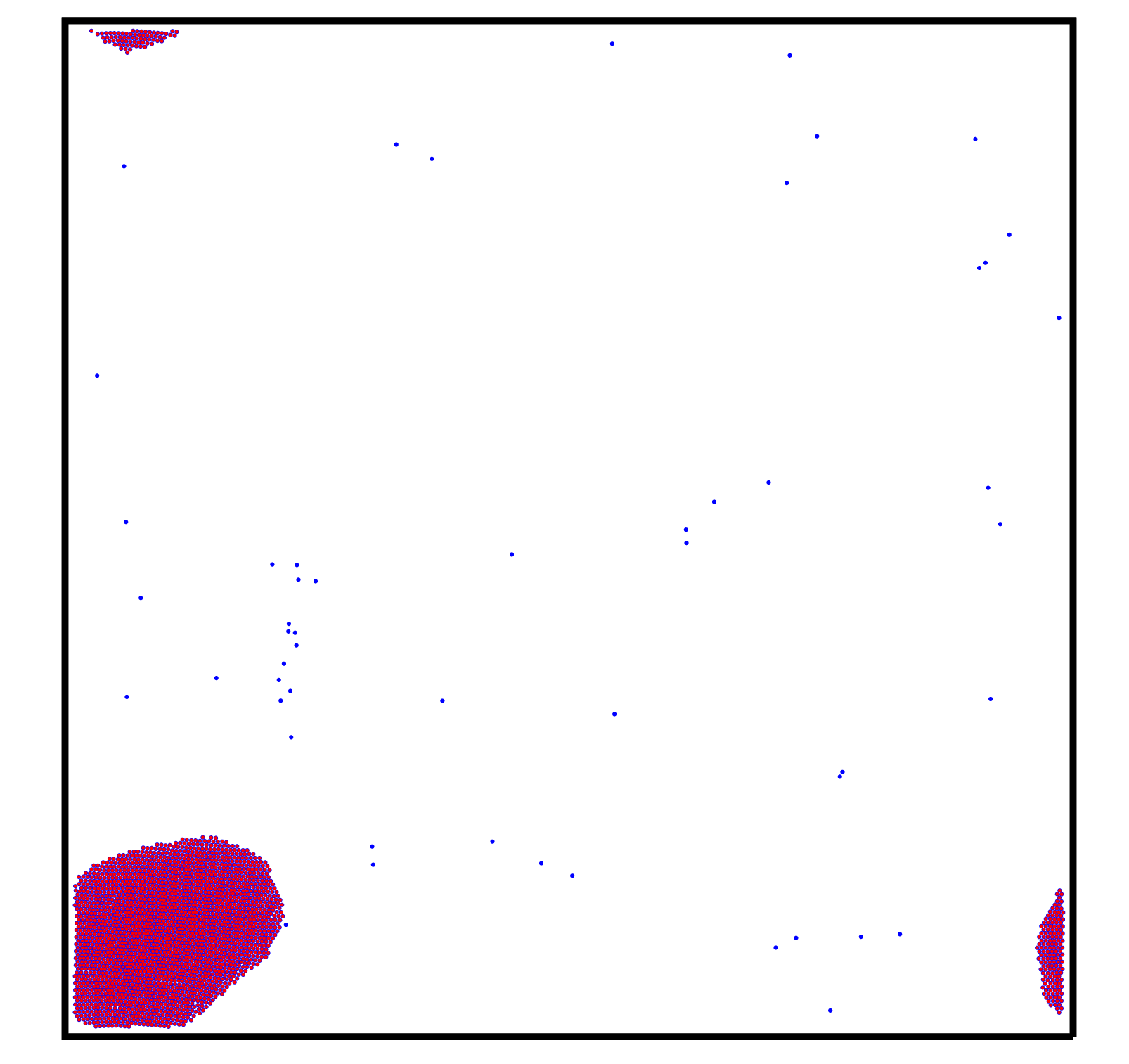,width=.32\linewidth} 
  \end{minipage} 
  \begin{minipage}{0.02\linewidth} 
    \begin{sideways} 
      $\lambda/L=10.0$ 
    \end{sideways} 
  \end{minipage} 
  \begin{minipage}{0.968\linewidth} 
    \epsfig{file=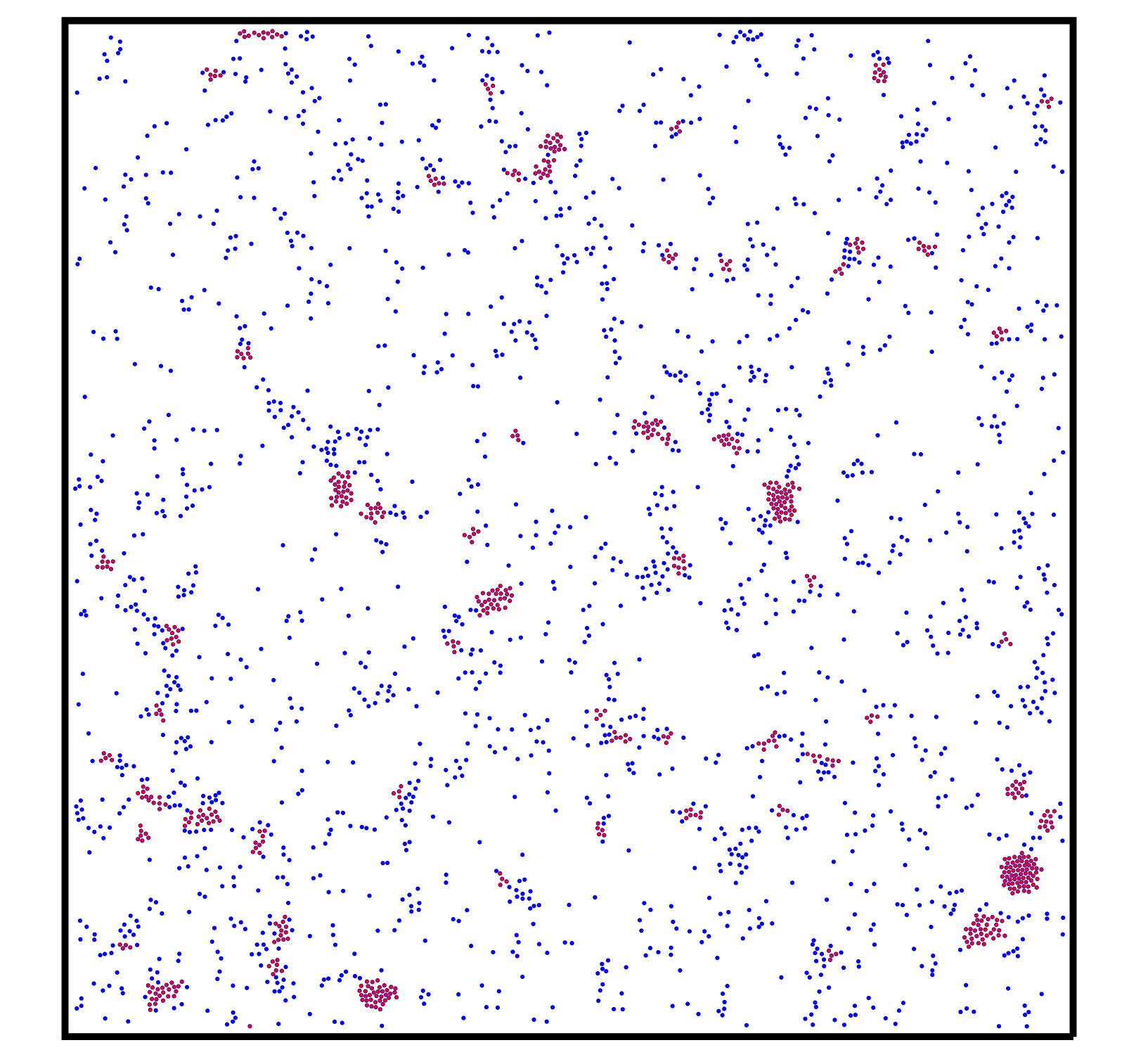,width=.32\linewidth} 
    \epsfig{file=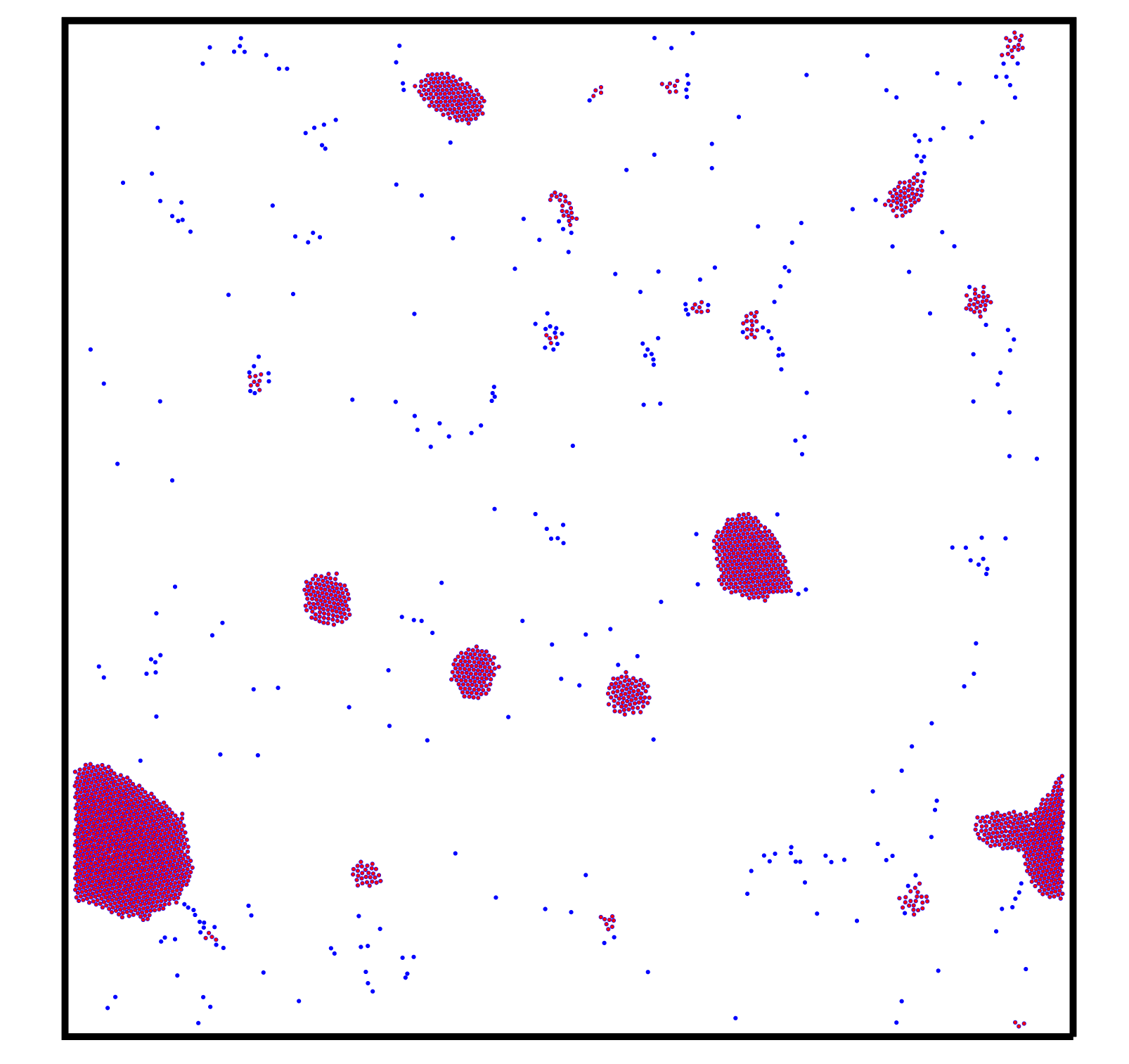,width=.32\linewidth} 
    \epsfig{file=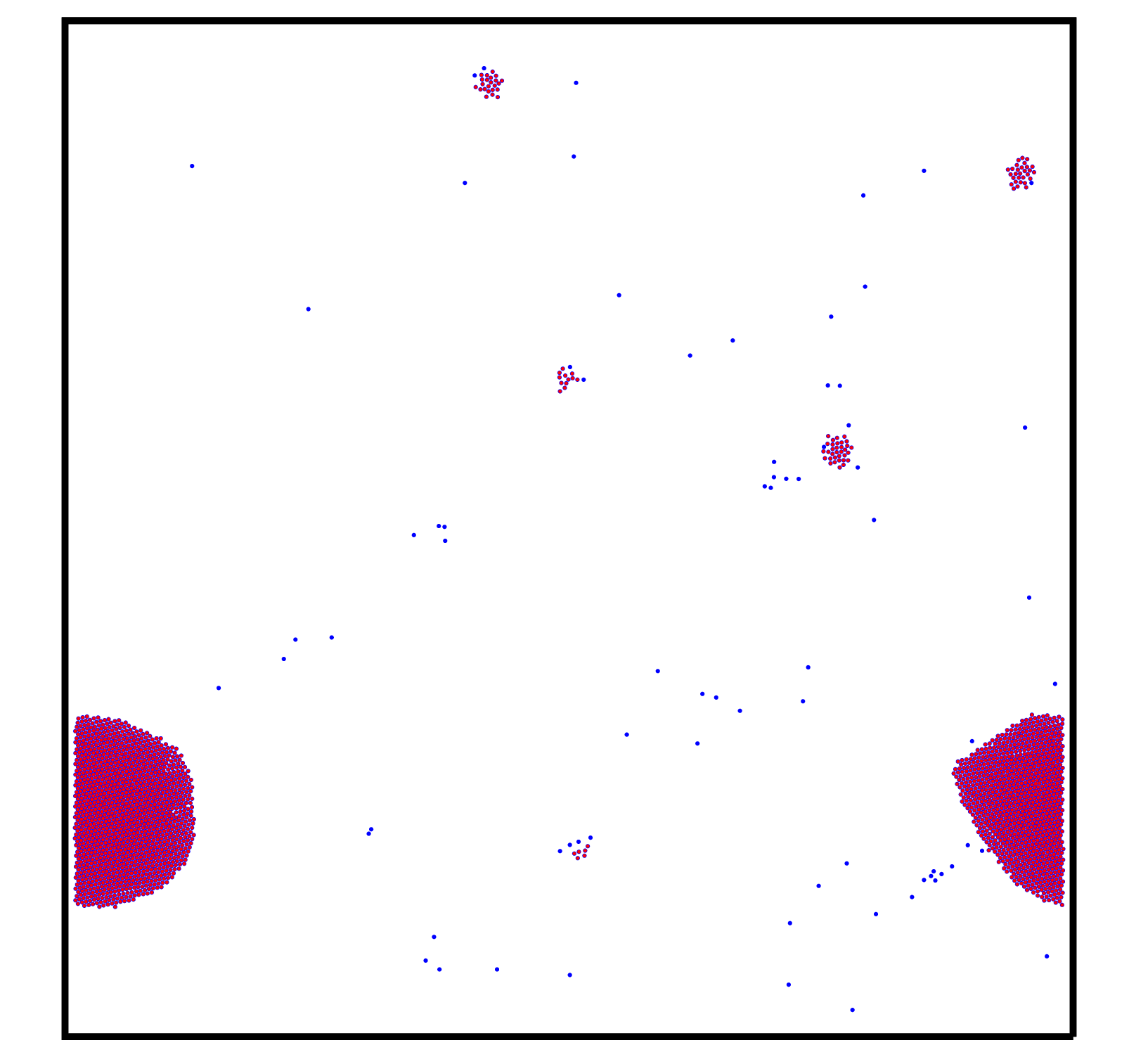,width=.32\linewidth} 
  \end{minipage} 
  \caption{\label{fig_6}The same presentation as in Fig.~\ref{fig_5}  
    for a ``long--ranged'' capillary 
    attraction with $\lambda/L=0.54$, $\lambda/L=3.0$, and $\lambda/L=10.0$ 
    (from top to bottom). } 
\end{figure} 
 
The simulations allow us to investigate the highly nonlinear evolution 
following the initial clustering stage and  
thus complement the theoretical analysis presented 
in Ref.~\cite{Dominguez:2010}. 
In Figs.~\ref{fig_5} and \ref{fig_6} we show the particle distribution in real 
space at time steps between $\mathcal{T}$ and $5 \mathcal{T}$ for several 
values of $\lambda$, while keeping the strength $V_0$ of the capillary 
potential fixed (see Eq.~(\ref{eq2})). We define a particle to be a member 
of a cluster (and thus being colored in red) if there are at least three 
neighboring particles at a distance  
$d < 3.25 R_0$. This choice is motivated by the position $r_{min} \approx 3.25 
R_0$ of the first minimum in the pair correlation function for a dense packing 
of discs with the repulsion given by Eq.~(\ref{eq:WCA}), as deduced from 
simulations.   
For $\lambda \simeq L$ (see Fig.~\ref{fig_6}) the system exhibits the formation 
of a few clusters on the time scale set by Jeans' characteristic time 
$\mathcal{T}$. These then merge quickly leaving the system in its final state 
of a single cluster, which contains basically all colloids. For the longest 
range of the capillary potential probed by us, $\lambda/L=10$, almost all 
particles in the  
system collapse into a single cluster between $t=3 \mathcal{T}$ and $t=4 
\mathcal{T}$. These observations provide a qualitative confirmation of the 
prediction according to  Eq.~(\ref{eq:Tcoll}) that in the Newtonian limit of the 
cold--collapse approximation the time scale of evolution is of the order of 
Jeans' time.  
 
\begin{figure}[ht] 
  \epsfig{file=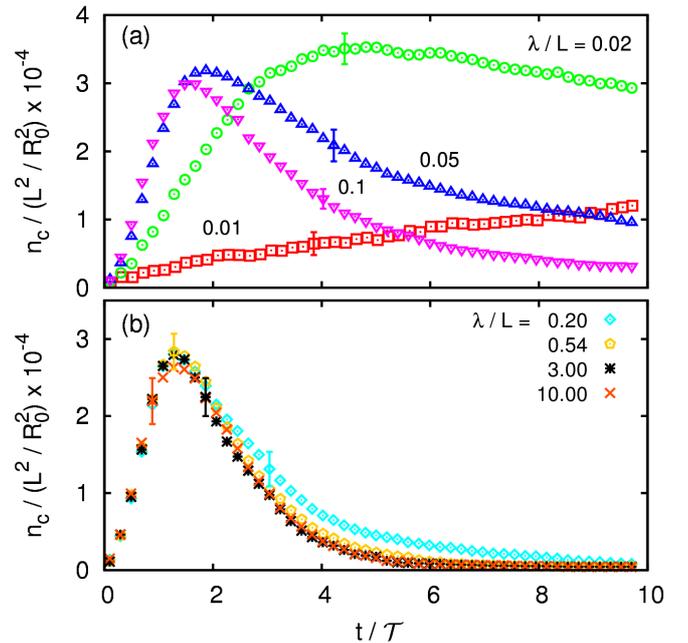,height=\linewidth,angle=270} 
  \caption{\label{fig_7}Evolution of the areal number density 
    $n_\mathrm{c}/L^2$ of clusters for several values of $\lambda/L$, 
    exhibiting the crossover from ``short--ranged'' attraction 
    (upper plot (a), $\lambda \lesssim L$) to ``long--ranged'' attraction  
    (lower plot (b), $\lambda 
    \gtrsim L$). Only representative error bars are shown.} 
\end{figure}

\begin{figure}[ht] 
  \epsfig{file=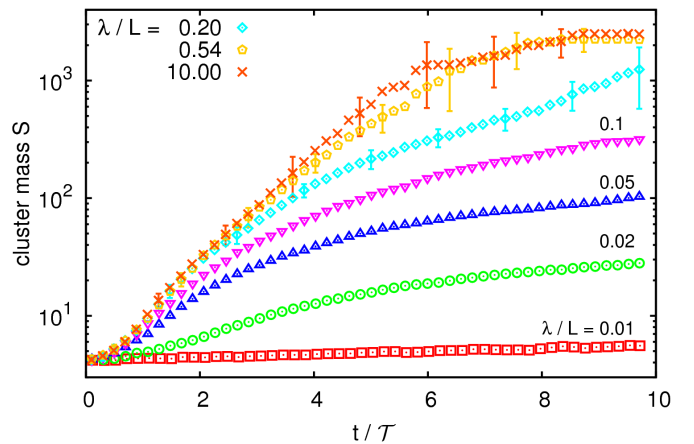,height=\linewidth,angle=270} 
  \caption{\label{fig_8}Evolution of the mean cluster mass $S(t)$ 
    (Eq.~(\ref{eq:mean_S})) for the same values of $\lambda$ as in 
    Fig.~\ref{fig_7}. Only representative error bars are shown. For $\lambda/L 
    \leq 0.1$ the error bars are of the size of the symbols.} 
\end{figure}

In order to obtain a more quantitative description  
of these clustering phenomena, the 
temporal evolution of the total number of clusters $n_\mathrm{c}$  
(regardless of their size) is presented in Fig.~\ref{fig_7}, while 
Fig.~\ref{fig_8} shows the   
mean cluster mass defined as~\cite{Meakin:1992}   
\begin{equation} 
  \label{eq:mean_S} 
   S(t)=\frac{\sum_{s=1}^N s\, n_\mathrm{c}(s,t)}{\sum_{s=1}^N 
     n_\mathrm{c}(s,t)}\quad,  
\end{equation} 
where $n_\mathrm{c}(s,t)$ is the number of clusters at time $t$ which consist 
of $s$ particles (note that $n_\mathrm{c}(s\leq 3,t)=0$ according to the 
definition of clusters introduced before).  
%%%%%%%%%%%%% 
As $\lambda$ increases, the evolution of these 
observables converges to a $\lambda$--independent limit characterized 
by scaling time $t$ with Jeans' time. This can be identified with the Newtonian 
limit ($\lambda/L \to \infty$) and with the corresponding onset of  
``long--ranged'' dynamics analogous to the gravitational collapse known 
from astrophysical scenarios. 
In this limit the temporal evolution extends over a rather short time 
period ($\sim 5 \mathcal{T}$), but nevertheless one can interpret 
Fig.~\ref{fig_7} qualitatively in terms of a two--stage aggregation 
process: The evolution is initially dominated by the formation of new 
clusters up to a crossover time $\sim 1.5 \mathcal{T}$, after which the 
evolution is dominated by the merging of clusters. 
This kind of process is not borne out so obviously by Fig.~\ref{fig_8} which 
confirms the steady increase of the average cluster mass, and which seems to 
follow an exponential law before saturation, induced by finite--size effects, 
sets in.  
This latter observation suggests that the average mass $S(t)$ is biased 
towards the largest cluster, so that the early evolution of the mass 
$s_\mathrm{large}(t)$ of the largest cluster can be described by a simple 
model of accretion of particles (either isolated or in the form of small 
clusters) at a rate proportional to the capillary attraction by the 
large cluster, i.e., 
\begin{equation} 
  \frac{d s_\mathrm{large}}{d t} = k s_\mathrm{large} ,  
  \qquad 
  k=\mathrm{const.} 
\end{equation} 
 
As a complement to these observables,  
we have also determined the Minkowski functionals (also known as 
intrinsic volumes, quermassintegrale, or curvature integrals) 
associated with the density field. Inter alia, these quantities have been  
applied successfully to the pattern analysis of spinodal decomposition 
in a 2D van--der--Waals gas \cite{Mecke:1999}. They provide valuable 
morphological information about the structure formation which is barely 
accessible by other standard probes such as two--point correlation functions 
because the Minkowski functionals are sensitive to higher--order 
correlations.  
%%%%%%%%%%% 
The properties and applications of these quantities in 
statistical physics are reviewed in Ref.~\cite{Mecke:2000}. Here we provide 
only the basic corresponding informations required for our present 
purposes. For a given geometrical domain $\mathcal{D}$ in the 2D plane, we 
define the three possible Minkowski functionals $M_\mathrm{\nu}(\mathcal{D})$ 
as (there are differences in the literature concerning a conventional geometric 
prefactor)  
\begin{align} 
  M_0 (\mathcal{D}) & = \textrm{area of $\mathcal{D}$} \label{def:M0}, \\ 
  M_1 (\mathcal{D}) & = \textrm{length of the boundary of 
    $\mathcal{D}$} \label{def:M1} , \\ 
  M_2 (\mathcal{D}) & = \textrm{Euler characteristic of 
    $\mathcal{D}$} \label{def:M2}  . 
\end{align} 
The Euler characteristic $\chi$ is a measure of the domain 
connectivity~\cite{Mecke:1999,Mecke:2000}: 
\begin{equation} 
  \label{def:chi} 
  \chi = \textrm{number of clusters minus number of holes} , 
\end{equation} 
where here a ``cluster'' is defined as a connected subdomain of 
$\mathcal{D}$ not connected 
to other ``clusters'' (i.e., isolated), 
%%%%%%%%%%%% 
and a ``hole'' is defined as  
a ``cluster'' in the complement of $\mathcal{D}$, i.e., in the domain 
$\mathbb{R}^2 \backslash \mathcal{D}$.  Alternatively, $\chi$ can be computed 
also as the integral of the curvature along the boundary of $\mathcal{D}$ and 
thus provides a compact measure of whether the boundary of $\mathcal{D}$ is 
predominantly convex or concave. For a given density field $\varrho(\br)$ and 
chosen threshold values $\varrho_\mathrm{c}$, one can construct domains 
$\mathcal{D}(\varrho_\mathrm{c})$ as the regions where the density is larger 
than the threshold value:  
\begin{equation} 
  \mathcal{D}(\varrho_\mathrm{c})=\left\{ \br \, | \, \varrho(\br) > 
  \varrho_\mathrm{c} \right\} .  
\end{equation} 
The density field $\varrho(\br)$ is computed on a grid with a 
mass--assignment scheme as described in Sec.~\ref{sec:simulations}, except 
that the grid is coarser than the one used to compute the capillary force (for 
which the grid spacing is  $r_\mathrm{c} = R_0$, see Table~\ref{tab1}).  
%%%%%%%%%%%% 
The purpose of this choice is to enhance the smoothening of the density grid; 
the chosen grid spacing of $r_\mathrm{Mink.} = 5.3 R_0$ offers an acceptable 
compromise between smoothness, accuracy, and computational cost.  
%%%%%%%%%%%%    
Accordingly, the density field subjected to such a threshold reduces to a 
pattern described by pixels and the domain $\mathcal{D}(\varrho_\mathrm{c})$ 
is the set of, say, white pixels, for which the Minkowski functionals can be 
computed efficiently (for details see Ref.~\cite{Mecke:1999}). 
 
\begin{figure}[ht] 
  \epsfig{file=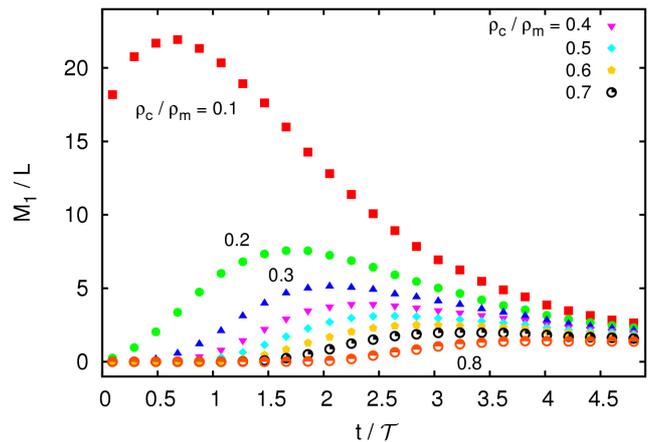,height=\linewidth,angle=270} 
  \caption{\label{figM1_rhoc}Evolution of $M_1$ (see Eq. (\ref{def:M1})) for 
    various values $\varrho_c$ of the density threshold within the range $0.1 
    \leq \varrho_\mathrm{c}/\varrho_\mathrm{m} \leq 0.8$, which corresponds to the 
    range $3 \leq \varrho_\mathrm{c}/\varrho_\mathrm{h} \leq 23$, for  
    $\lambda/L= 0.54$, $\varrho_\mathrm{m}=29 \varrho_\mathrm{h}$, and 
    $\varrho_\mathrm{h} = 10 R_o^{-2}$. Note that for large times, $M_1 /
    L$ converges to the circumference of a close packed disc
    ($M_1/L=\frac{2\pi}{L}\sqrt{N/(\pi\varrho_\mathrm{m})}\approx
    0.66$; not yet visible on the present scale) for any density threshold.
    For $\varrho_\mathrm{c}/\varrho_\mathrm{m}=0.1$ the error bars 
    are twice as large as the symbol, in the other cases they are as large as
    or smaller than the symbol size. } 
\end{figure}

\begin{figure}[ht!] 
  \epsfig{file=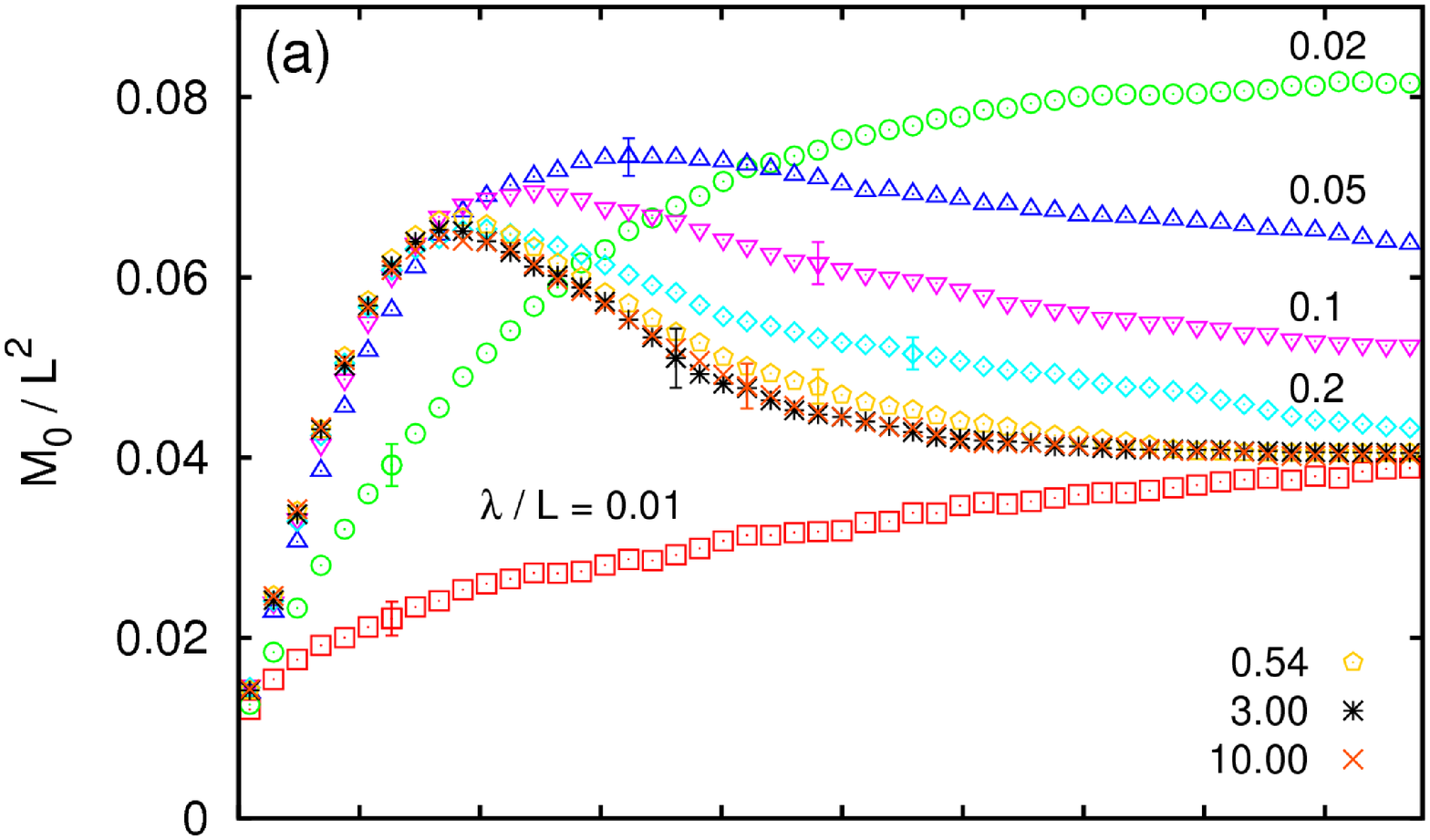,height=\linewidth,angle=270} 
  \epsfig{file=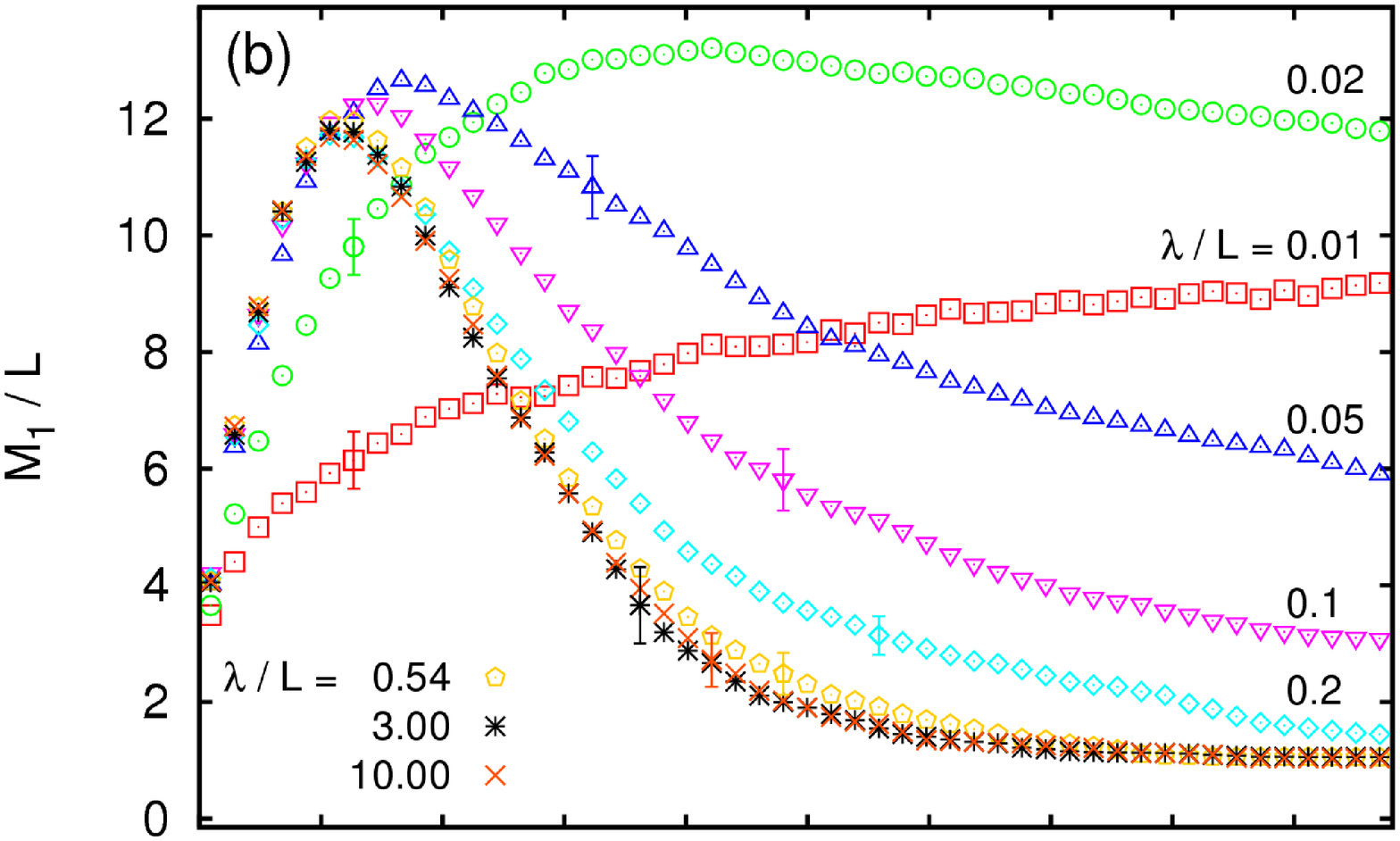,height=\linewidth,angle=270} 
  \epsfig{file=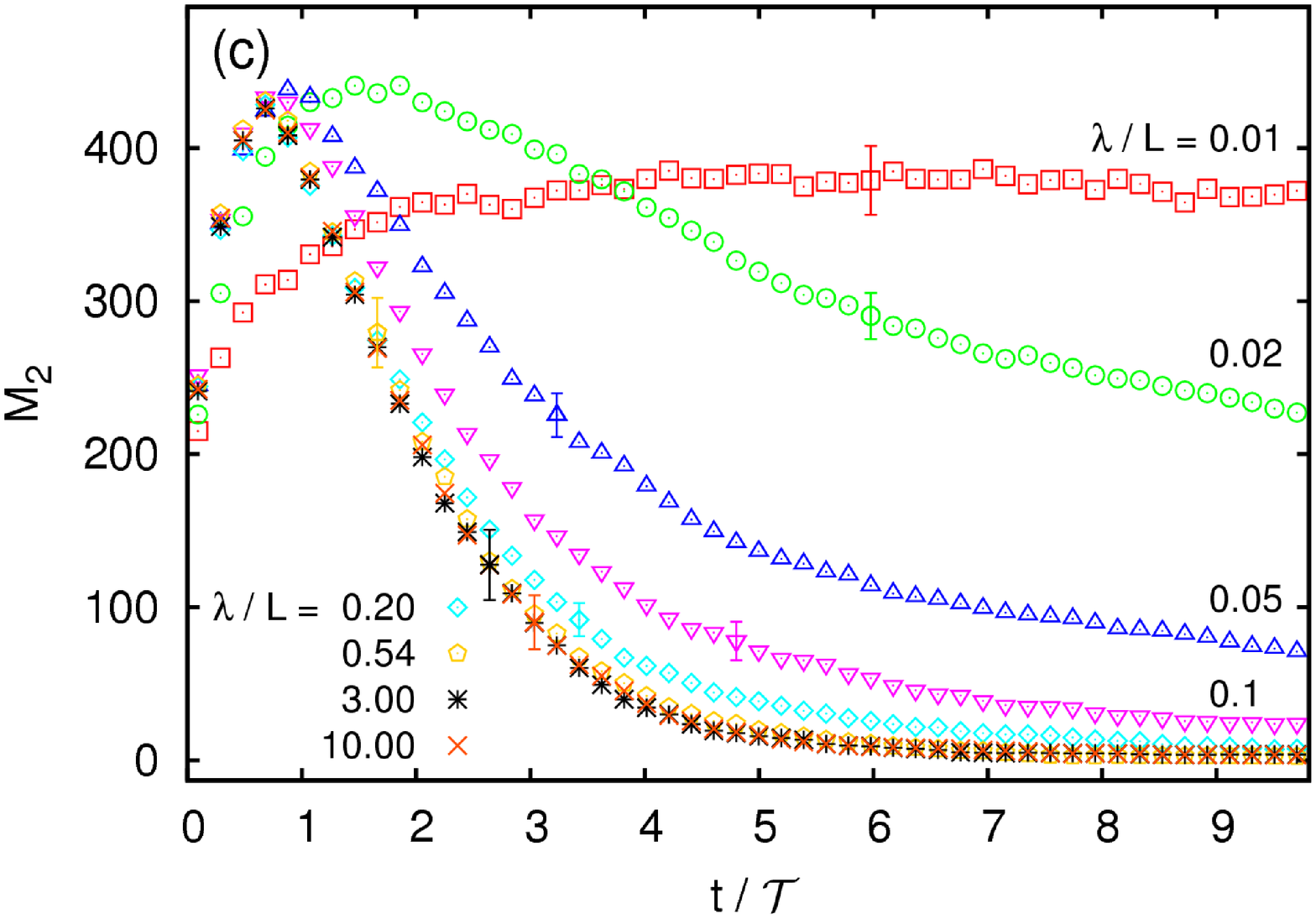,height=\linewidth,angle=270} 
  \caption{\label{figM0}Evolution of the Minkowski %functional 
    functionals $M_0$ (a), $M_1$ (b), and $M_2$ (c) 
    (see Eq.~(\ref{def:M0})) for the threshold $\varrho_\mathrm{c}=4 
    \varrho_\mathrm{h}  = 0.14 \varrho_\mathrm{m}$ with 
    $\varrho_\mathrm{h} = 10 R_0^{-2}$ and for the same values of $\lambda$ 
    considered in  
    Fig.~\ref{fig_7}. $M_0$ is proportional to the area of the region where 
    the density is higher than the threshold value  $\varrho_\mathrm{c}$. 
    $M_1$ is proportional to the  contour length of the boundary of this
    region whereas $M_2$ measures the corresponding Euler
    characteristic.    
    All 
    other parameters of the system are chosen as given in Table 
    \ref{tab1}. Only representative error bars are shown.}  
\end{figure}

\begin{figure}[ht] 
  \epsfig{file=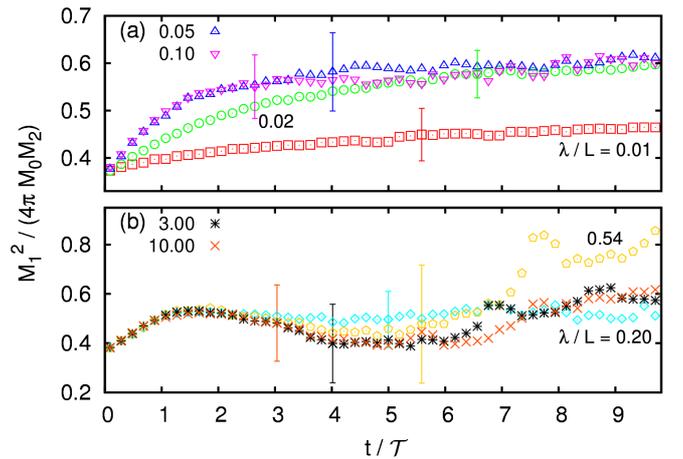,height=\linewidth,angle=270} 
  \caption{\label{fig:shape} Same as Fig.~\ref{figM0} for the quantity 
    $\phi = M_1^2/(4 \pi M_0 M_2)$ related to the typical 
    shape of clusters and for the threshold 
      $\varrho_\mathrm{c}=0.14\varrho_\mathrm{m}$.  
    Note the different scales in panels (a) and (b). Only representative error 
    bars are shown.}  
\end{figure}

By construction, the Minkowski functionals are functions of time and of the 
density threshold $\varrho_\mathrm{c}$. The latter dependence has been probed 
in the range between the initial number density $\varrho_\mathrm{h} = (10 
R_0)^{-2}$ and the density $\varrho_\mathrm{m} = 1/(2\sqrt{3} R_0^2) \approx 
29 \varrho_\mathrm{h}$ of close packing of discs of radius $R_0$. The 
qualitative behavior of the functionals does not depend on 
$\varrho_\mathrm{c}$ if this value falls within this range (an example for the 
dependence on the threshold value is presented in Fig.~\ref{figM1_rhoc}).   
Therefore, in Fig.~\ref{figM0} %--\ref{figM0} 
the time 
evolution of the three Minkowski functionals is shown only for the threshold 
value $\varrho_\mathrm{c} = 0.14 \varrho_\mathrm{m} \approx 4 
\varrho_\mathrm{h}$. 
If the threshold value is sufficiently large, as it is the case for 
$\varrho_\mathrm{c} = 4 \varrho_\mathrm{h}$, it is expected that in the domain 
$\mathcal{D}$ no ``holes'' appear, so that $M_2>0$ and gives directly the 
number of clusters. Furthermore, provided the system is ergodic, $M_0$ is 
proportional to the probability distribution of an overdensity, i.e., that the 
density at any point is larger than $\varrho_\mathrm{c}$.  
$M_1$ is a direct measure of the length of the 
``interface'' separating clusters (i.e., over--dense regions) from the 
surrounding low--density gas of particles. The quantities $M_0/M_2$, 
$M_1/M_2$ provide an estimate of the average cluster area and 
perimeter, respectively. We define the ``shape factor'' $\phi := 
M_1^2/(4 \pi M_0 M_2)$ which provides a partial, quantitative characterization 
of the distribution of cluster shapes (see Fig. \ref{fig:shape}). It is 
normalized such that $\phi=1$ for a single circular cluster.   
 
% 
%\begin{figure}[ht] 
%  \epsfig{file=fig-9.eps,height=\linewidth,angle=270} 
%  \caption{\label{figM1} Same as Fig.~\ref{figM0} but for the Minkowski 
%    functional $M_1$ and the threshold 
%      $\varrho_\mathrm{c}=0.14\varrho_\mathrm{m}$ (see 
%      Eq.~(\ref{def:M1})). This quantity is proportional to the  
%    contour length of the boundary of the region where the density is higher 
%    than the threshold value  $\varrho_\mathrm{c}$.  
%    Only representative error bars are shown.  } 
%\end{figure}  
% 
%\begin{figure}[ht] 
%  \epsfig{file=fig-10.eps,height=\linewidth,angle=270} 
%  \caption{\label{figM2} Same as Fig.~\ref{figM0} for the Minkowski
%    functional $M_2$ and the threshold 
%      $\varrho_\mathrm{c}=0.14\varrho_\mathrm{m}$ (see Eqs. (\ref{def:M2}) 
%    and (\ref{def:chi})). This quantity is  
%    proportional to the Euler characteristic of the region where the density 
%    is higher than the threshold value  $\varrho_\mathrm{c}$.  
%    Only representative error bars are shown.} 
%\end{figure}  
 
It is interesting to try to understand the simulation results within the 
framework of spinodal decomposition. In our case, the homogeneous initial 
distribution (which is the stable equilibrium state   
if $\lambda K < 1$, see Eqs. (\ref{eq:jeansK}) and (\ref{eq:Kc})) 
is destabilized by a quench into a coexistence region between a high--density, 
liquid phase ($\varrho \sim \varrho_\mathrm{m}$) and a rather dilute, 
gaseous phase. Since the average density $\varrho_\mathrm{h}$ is substantially 
smaller than $\varrho_\mathrm{m}$, this can be viewed as a quench off the 
center of the coexistence region with the liquid state being the minority 
phase. In this context one typically identifies two successive regimes: (i) 
the early stage of \textit{phase separation} when droplets (i.e., clusters) of 
the minority phase grow out of the initial fluctuation--induced seeds by 
binding individual particles, and (ii) the late stage of \textit{domain 
coarsening} when the droplets coalesce. In Ref.~\cite{Mecke:1999}, the 
occurrence of the crossover between these two stages has been identified based 
on the behavior of the position of the first maximum of the Minkowski 
functionals as function of time. 
 
Inspection of Figs.~\ref{fig_7}--\ref{fig:shape} suggests that this 
spinodal decomposition scenario is an appropriate interpretation of 
the present evolution. For $\lambda/L =0.01$, which is the smallest  
value of $\lambda$ studied here, one observes only the initial stage of phase 
separation, marked by a steady increase of $n_\mathrm{c}$ and $M_2$ 
(number of clusters), $S$ (average mass), $M_1$ (accumulated 
perimeter), and $M_0$ (overdensity area).  Apparently for this small 
value of $\lambda$ the crossover to domain coarsening occurs later 
than the maximum time of our simulations.  
However, as evidenced for somewhat larger values of $\lambda$, the 
shape factor $\phi$ tends to become time--independent, indicating a 
self--similar growth of the clusters in the domain--coarsening stage. 
In the opposite limit, for the largest values of $\lambda$ explored 
the crossover to the second stage is clearly observable. It is 
characterized by the occurrence of maxima followed by a steady decrease of the 
number of clusters, 
of the accumulated perimeter, of the overdensity area, and of the 
shape factor. After the system has collapsed to a single cluster at time 
$t\approx 6 \mathcal{T}$, the increase of the shape factor $\phi$, although 
subject to fluctuations, indicates a further compactification and 
reorganization of the clustered particles towards forming a circular shape.   
The average mass $S$, however, does not exhibit any 
clear feature indicative of the crossover. It eventually saturates 
when only a single cluster remains, to which most particles are bound. 

We emphasize that this description is valid if the threshold density 
$\varrho_\mathrm{c}$ takes intermediate values. If 
$\varrho_\mathrm{c}$ is close to the maximum density 
$\varrho_\mathrm{m}$, the signature of the crossover is no longer 
detectable (see Fig.~\ref{figM1_rhoc}), because in this case the 
Minkowski functionals just probe the very dense regions, which are 
formed only at late times and for which the dynamics is presumably no 
longer dominated by capillary attraction but is instead strongly 
affected by the short--ranged repulsion.  
 
A characteristic feature of the coarsening stage 
during spinodal decomposition of a common fluid is a self--similar 
dynamical scaling. 
As we have seen, if  
$\lambda$ is large the evolution is too fast and the range of times 
probed by the simulations (a few times $\mathcal{T}$) is too narrow in 
order to be able to reliably observe a self--similar evolution. 
However, one can still identify the coarsening stage by means of the 
extrema in the Minkowski functionals. This demonstrates their 
robustness even in the extreme case of the almost simultaneous 
collapse of all length scales, i.e., if most of the wavevectors of   
the system fulfill $k<K_{\rm c}$ (see Eq. (\ref{eq:Kc})) so that the 
corresponding Fourier components grow rapidly. This 
supports the applicability of the Minkowski functionals as advocated in 
Ref.~\cite{Mecke:1999}.  
 
In the absence of an elaborated theory, here we can provide only a 
tentative explanation concerning specifically the influence of 
$\lambda$ on the evolution. Our numerical observations can be 
rationalized assuming that the dependence on $\lambda$ enters only via 
the two dimensionless quantities $\lambda K$ and $\lambda/L$. The data 
suggest that the Newtonian limit ($1 \ll \lambda/L$, $\lambda K$) 
yields a $\lambda$--independent behavior and that the 
crossover occurs at a time of the order of Jeans' time $\mathcal{T}$, 
in agreement with Eq.~(\ref{eq:Tcoll}). For intermediate values of 
$\lambda$ ($\lambda/L \ll 1 \ll \lambda K$), only the early--time 
evolution is $\lambda$--independent (in agreement with the analytic 
linear analysis, see Eq.~(\ref{eq:tau})) whereas the late--time, 
domain--coarsening stage depends significantly on $\lambda$, but the 
crossover time does so only weakly. Finally, the whole evolution 
depends sensitively on the value of $\lambda$ if it is small 
($\lambda/L \ll \lambda K \sim 1$): for $\lambda \gtrsim 1/K$, the homogeneous 
state is unstable and undergoes spinodal decomposition, whereas for $\lambda 
\lesssim 1/K$ the homogeneous state becomes stable as estimated from mean--field 
theory (see Sec.~\ref{sec:theory}).    
 
\subsection{Radially symmetric initial configuration} 
\label{4} 
\begin{figure}[ht!] 
  \epsfig{file=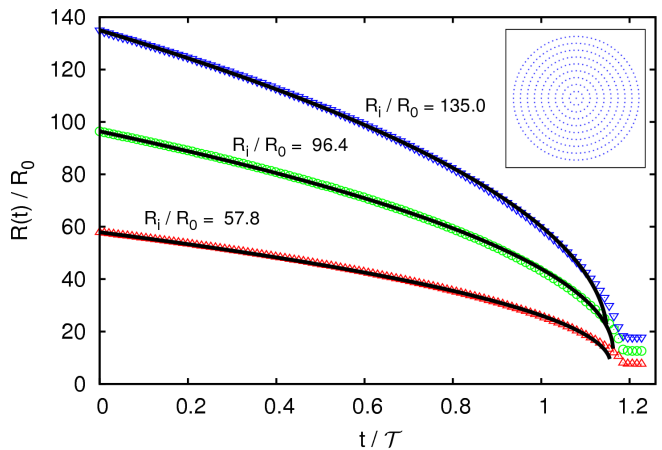,height=\linewidth,angle=270} 
  \caption{\label{fig_13} 
    Cold evolution of the radii $R(t)$ of three
    rings with initial radius $R_\mathrm{i}/R_0
    =R(t=0)/R_0 =58,\,95,\,\textrm{and}\,135$ for the highly
    ordered initial configuration of concentric rings forming a disc as 
    discussed in the main text (see inset). The full lines correspond to 
    Eq.~(\ref{eq:real_radius}) with 
    $(A_0,\,R_\mathrm{f}/R_0)=(1.15,\,18),\,(1.16,\,13),\,\textrm{and}
    \,(1.16,\,8)$ top down. 
    The quadratic simulation box of size 
    $L^{\prime} = 800 R_0$ contains initially a disc of radius 
    $L= 183 R_0$. The number density of particles inside the 
    disc is $\varrho_\mathrm{h} R_0^2=4.28\times 10^{-3}$ 
    The range of the capillary attraction is $\lambda/L=1.48$ and
    $\mathcal{T}=95340$ s.}
\end{figure}

In this subsection we test the cold--collapse approximation for radially 
symmetric configurations. In this context ``cold'' means that the particles 
move under the action of the capillary force only, corresponding to the formal 
limits  
$T\to 0$ (no Brownian motion) and $R_0\to 0$ (no short--ranged repulsion). 
The first limit is easily attained by switching off 
the random displacement i.e., $\mathring{\br}=0$ in Eq.~(\ref{eq3}) leading to 
a deterministic Langevin equation (see Eq.~(\ref{eq:width})).  
In order to comply with the 
second limit we adopt a sufficiently large initial mean interparticle 
separation, so that the cold--collapse evolution can be followed numerically 
over some time. This way the breakdown of the cold--collapse approximation can 
be monitored realistically in those regions where the density approaches its 
largest possible value.  
All simulations have been carried out for $N=451$ particles 
 arranged as a disc with initial radius $L = 183 
R_0$ contained in a quadratic simulation box with side length $L^{\prime}=800 
R_0$.   
Radial symmetry is achieved approximately by surrounding %as before  
the collapsing disc with a ringlike stripe of vacuum of thickness 
$\approx 220 R_0$.   
%%%%%%%%%%% 
For simulations in which the radius of the disc is smaller than  
$\lambda$, 
the additional stripe of vacuum ensures  
that the periodic images, which violate the rotational symmetry, contribute 
only weakly to the total force experienced by the particles.    
%%%%%%%%%%% 
We have run two sets of simulations: 
\begin{enumerate} 
\item In the first set, temperature has been switched off and
  particles are initially ordered in suitable concentric rings with
  equal spacing (arc length $\approx 12.12 R_0$) along the
  circumference of all rings, so that each ring possesses discrete
  azimuthal invariance\footnote{Each ring is azimuthally invariant
    with respect to multiples of the angular separation of nearest
    neighbors in the same ring.} (see inset of
  Fig.~\ref{fig_13}). The azimuthal position of the particles is
  preserved during evolution.  These somewhat artificial conditions
  provide a way to estimate the influence of the finite values of
  $R_0$ and $\lambda$ on the evolution as compared to the theoretical
  prediction, summarized in Eq.~(\ref{eq:radius}).
\item In the second set the previous conditions are relaxed: 
  temperature is switched on as it has been done in the simulations 
  discussed in the   
  previous subsection, and the initial particle 
  distribution inside the disk is random. With this more realistic 
  study we address the effect nonzero temperature has on the 
  theoretical prediction. 
\end{enumerate} 
 
\begin{figure}[ht] 
  \epsfig{file=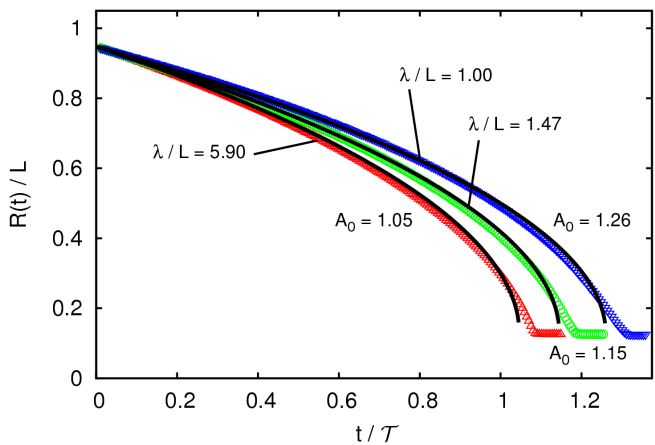,height=\linewidth,angle=270} 
  \caption{\label{fig_14}Same as Fig.~\ref{fig_13} for the outermost
    ring ($R_\mathrm{i}/R_0 = 174$, which is smaller than $L$ due to
    binning) for $\lambda/L = 1,\,1.48$, $5.9$ and initial disc radius
    $L=183R_0$.  The full lines correspond to the expression in
    Eq.~(\ref{eq:real_radius}). As expected from Eq.~(\ref{eq:radius})
    it turns out that $A_0(\lambda \to \infty)=1$.}
\end{figure}

For the first set of simulations, Fig.~\ref{fig_13} 
shows the evolution of the radial coordinate $R(t)$ for three rings inside 
this disc with different initial radii $R_\mathrm{i}$. Motivated by the 
theoretical prediction in Eq.~(\ref{eq:radius}), which has been obtained in 
the Newtonian limit within the cold--collapse approximation,  
the time dependence can be fitted rather well by the following expression:  
\begin{equation} 
  \label{eq:real_radius} 
  R(t) - R_\mathrm{f} = \left( R_\mathrm{i} - R_\mathrm{f} \right)  
  \sqrt{1 - \frac{t}{A_0 \mathcal{T}}} . 
\end{equation} 
This expression differs from  
Eq.~(\ref{eq:radius})  
 in two respects: the ring radius $R(t)$ is shifted by 
subtracting the final radius $R_\mathrm{f}$, and the time scale $\mathcal{T}$ 
of collapse is dilated by a factor $A_0$.  
Deviations at later times, such as the smearing out of the square root 
singularity, can be clearly attributed to the non--vanishing 
size of the particles and the resulting repulsive forces. 
The effect of the finite 
compressibility comes into play in the late stage of the evolution  
and prevents the vanishing of $R(t)$, so that 
$R_\mathrm{f} \neq 0$.  
%%%%%%%%%%%% 
The factor $A_0 \approx 1.16$ follows from a fit in the range $0 < t <
0.9 \mathcal{T}$. However, as can be seen in Fig.~\ref{fig_13}, the
time $A_0\mathcal{T}$ offers a reasonable estimate for the
\textit{effective} time of collapse in the simulations
$\mathcal{T}_\mathrm{coll} \approx A_0\mathcal{T}$, defined by the
  condition $R(\mathcal{T}_\mathrm{coll})=R_\mathrm{f}$.
%%%%%%%%%%%% 
It turns out that the factor $A_0$   
depends only weakly on $R_\mathrm{i}$ and we conjecture that the 
delayed effective time of collapse $\mathcal{T}_\mathrm{coll}$, i.e., $A_0 >1$, 
is a consequence of the finite value of 
$\lambda$.   
Additionally, residual effects of the tidal forces of the periodic 
images might be non--negligible. Both effects tend to weaken the attractive 
force of the collapsing cluster and thus increase $\mathcal{T}_{\rm coll}$.   
In order to support this conjecture, in Fig.~\ref{fig_14} we show the 
evolution  
of the radius $R(t)$ of the outermost ring (actually $R_i/R_0 < L$ due 
to binning and the applied discretization scheme) 
for three values of the screening length $\lambda$. 
Upon increasing $\lambda$ the collapse time  
decreases, eventually leading to a collapse at 
$\mathcal{T}_\mathrm{coll}=\mathcal{T}$, i.e.,  
$A_0(\lambda \to \infty)=1$ in line with Eq.~(\ref{eq:radius}).   
The delay in the effective time of collapse
$\mathcal{T}_\mathrm{coll}$ upon decreasing $\lambda$ is confirmed
in addition by a theoretical calculation of the cold--collapse
radial trajectories, which is perturbative in the parameter $1/\lambda$ (see
Eqs.~(4) and (5) in Ref.~\cite{Bleibel:2011}).  This effective time must
be understood as a fitting parameter (according to
Eq.~(\ref{eq:real_radius})) because the analysis presented in
Ref.~\cite{Bleibel:2011} indicates that for a small but non
vanishing value of $1/\lambda$ the (delayed) collapse does not
actually occur at the center but at the outer rim of the disk (forming
a shock wave within the cold--collapse approximation).

\begin{figure}[ht]  
  \epsfig{file=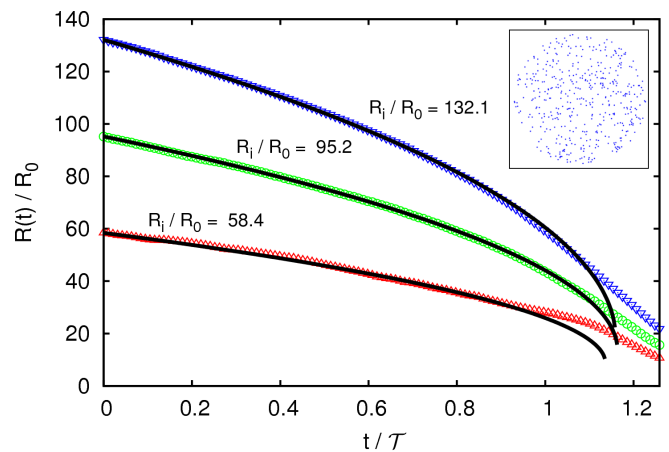,height=\linewidth,angle=270} 
  \caption{\label{fig_15} Same as Fig.~\ref{fig_13}, but for an initial random
    distribution of particles (see inset) inside a disc of initial radius 
    $R(0)/R_0=R_i/R_0$    
    and with Brownian motion switched on. The radial positions of the 
    particles have   
    been averaged over shells with a bin width of $\Delta r = 7.5 
    R_0$. The full lines correspond to Eq.~(\ref{eq:real_radius}) with 
    $(A_0,\,R_\mathrm{f}/R_0)=(1.16,\,18),\,(1.16,\,13),\,  
    \textrm{and}\,(1.14,\,9)$ from top to bottom.} 
\end{figure}

For the second set of simulations, 
Fig.~\ref{fig_15} presents the results after averaging over the data 
sets of ten runs corresponding to different realizations of the random 
initial configuration. Equation~(\ref{eq:real_radius}) 
provides a good fit and the conclusions following from it hold also for this 
kind of data set. 
The main differences with Fig.~\ref{fig_13} are that 
$R(t)$ approaches more smoothly its value $R_\mathrm{f}$ of the final, 
collapsed state, and that this happens at later times: the fit of the 
data for the same range $0 < t < 0.9\mathcal{T}$ 
shows much larger deviations at later times so 
that, although for a fixed value of $\lambda$ an approximately common fitting 
parameter $A_0$ can be found, $A_0\mathcal{T}$ does not provide an equally 
good estimate for the effective time of collapse $\mathcal{T}_\mathrm{coll}$.  
These deviations can be attributed to the Brownian 
motion and the random initial particle distribution. Both effects tend 
to spoil the radial symmetry and the sole prevalence of capillary 
attraction in driving the evolution and slow down the approach to the 
final state.  
For the purpose of comparison, the values for Jeans' time and wavenumber 
associated with the particle density $\varrho_\mathrm{h} = N/(\pi 
L^2)$ of the discs are $\mathcal{T}=95340\,$s and 
$K=15.2\,\mu$m$^{-1}$, respectively. Since $K^{-1} \ll L, L'$, one may 
have anticipated that the deviations induced by temperature 
are limited to the smallest length scale, i.e., $R_f$,  and relevant only after 
times of order $\mathcal{T}$, as observed indeed. 
 
\section{Summary and conclusions} 
\label{sec:summary} 
 
By means of numerical simulations, we have investigated the clustering 
of colloidal particles trapped at a fluid interface due to capillary 
attraction.  
As explained in detail in Sec.~\ref{sec:simulations}, our numerical 
approach deals with this long--ranged attraction by using a particle--mesh 
method as  
employed in simulations of self--gravitating fluids. This is motivated 
by the formal analogy between capillary and gravitational attraction 
and because each particle interacts simultaneously with many neighbors. 
The dynamics is Brownian and describes the overdamped motion of particles 
in fluids. 
The key parameters characterizing the macroscopic evolution of initial 
particle configurations are the 
range $\lambda$ of the capillary attraction (Eq. (\ref{eq2})), Jeans' length 
$1/K$ (Eq. (\ref{eq:jeansK})), and Jeans' time, $\mathcal{T}$ 
(Eq. (\ref{eq:jeansT})).

A distribution of particles is unstable against clustering provided 
the capillary attraction is, in a well--char-acterized 
sense (see Sec.~\ref{sec:theory}), sufficiently large. Our results confirm the 
predictions  
derived theoretically in Ref.~\cite{Dominguez:2010}: the dynamics evolve 
linearly during the early stage of evolution up 
to a time $\approx 1.5 \mathcal{T}$ (see Fig.~\ref{fig_2}). Also the 
nonlinear, radially symmetric collapse in the ``cold'' 
($K\to\infty$) as well as in the ``Newtonian'' ($\lambda\to\infty$) limit is 
well reproduced (see Figs.~\ref{fig_13}--\ref{fig_15}). There are small  
deviations from the theoretical predictions which are due to not reaching this 
formal double limit. 
 
In addition, the simulations provide insights which go beyond the simplifying 
approximations underlying the analytic theory. In particular, we have 
addressed the effect on the instability of the ratio $\lambda/L$ 
between the interaction range $\lambda$ and the system size $L$. 
We have analyzed the nonlinear evolution by using quantitative 
characteristics which are particularly sensitive to the process of structure 
formation. These are the number and the mass of clusters as well as the 
Minkowski  
functionals of the density field, which have been studied as function of time 
and of $\lambda/L$ (see Figs.~\ref{fig_7}--\ref{figM0}). 
In the ``short--ranged limit'' $\lambda\ll L$, the evolution 
proceeds  
on time scales much larger than $\mathcal{T}$. During the time probed 
in our simulations, the evolution is dominated by the steady formation 
of new clusters out of the seeds in the fluctuating density field. 
In the opposite, ``Newtonian'' limit $\lambda \gg L$, a  
$\lambda$--independent behavior emerges. The collapse proceeds much 
faster and is completed at a time $\approx 5\mathcal{T}$, thus confirming 
the conjecture put forward in Ref.~\cite{Dominguez:2010} that in this limit 
Jeans' time  
sets the temporal scale of the evolution. Despite the speed of this 
evolution, in the clustering process two regimes can be identified: an initial 
stage dominated by the formation of new clusters 
up to a time $\approx 1.5\mathcal{T}$, and a later stage dominated by 
cluster merging towards a single, large cluster. This is reminiscent 
of the spinodal decomposition dynamics in common fluids corresponding to 
the regimes of phase separation and domain coarsening, respectively. 
 
We interpret this $\lambda$--dependence of the dynamics as a crossover 
in the behavior from that of a 2D common fluid for small $\lambda$ (slow 
spinodal decomposition) to that of a self--gravitating fluid for large 
$\lambda$ (fast collapse). Our simulation results show to which extent the 
features of each limiting case is preserved in the other case and they suggest 
that the dependence on $\lambda$ is smooth. 
 
Together with the quantitative estimates for $\mathcal{T}$ and $K$ 
provided in Ref.~\cite{Dominguez:2010}, the present results are expected to be 
helpful for analyzing and interpreting corresponding experimental data. The 
simulation algorithm can be expanded easily to include the effect of external 
fields.  
 
%\begin{acknowledgments} 
\label{acknowl} 
J.B. and M.O. thank the German Research Foundation (DFG) for the financial 
support through the Collaborative Research Center (SFB-TR6) ``Colloids in 
External Fields'' Project N01. A.D.~acknowledges support by the Spanish 
Government  
through grants FIS2008-01339 (partially 
financed by FEDER funds) and AIB2010DE-00263. 
 
%\end{acknowledgments} 
 
%%%%%%%%%%%%%%%%%%%%%%%%%%%%%%%%%%%%%%%%%%%%%%%%%%%%%%%%%%%%%%%% 
 
%%%%%%%%%%%%%%%%%%%%%%%%%%%%%%%%%%%%%%%%%%%%%%%%%%%%%%%%%%%%%%%% 
 
\end{document}